%% file: 2404-020-3H-MasamitsuMori.tex
\documentclass[preprint]{ptephy_v1}

\preprintnumber{XXXX-XXXX} 

\usepackage{siunitx}
\usepackage{booktabs}

\usepackage{lineno}

\begin{document}

\title{Development of a data overflow protection system for Super-Kamiokande to maximize data from nearby supernovae}
\input{authors-20221203}


\begin{abstract}
Neutrinos from very nearby supernovae, such as Betelgeuse, are expected to generate more than ten million events over 10\,s in Super-Kamokande (SK). 
At such large event rates, the buffers of the SK analog-to-digital conversion board (QBEE) will overflow, causing random loss of data that is 
critical for understanding the dynamics of the supernova explosion mechanism. 
In order to solve this problem, two new DAQ modules were developed to aid in the observation of very nearby supernovae.
The first of these, the SN module, is designed to save only the number of hit PMTs during a supernova burst and the second, the Veto module, prescales the high rate neutrino events to prevent the QBEE from overflowing based on information from the SN module. 
In the event of a very nearby supernova, these modules allow SK to reconstruct the time evolution of the neutrino event rate from beginning to end using both QBEE and SN module data. 
This paper presents the development and testing of these modules together with an analysis of supernova-like data generated with a flashing laser diode. 
We demonstrate that the Veto module successfully prevents DAQ overflows for Betelgeuse-like supernovae as well as the long-term stability of the new 
modules. 
During normal running the Veto module is found to issue DAQ vetos a few times per month resulting in a total dead time less than 1\,ms, and does not influence ordinary operations.
Additionally, using simulation data we find that supernovae closer than 800~pc will trigger Veto module resulting in a prescaling of the observed neutrino data.
\end{abstract}

\subjectindex{xxxx, xxx}

\maketitle

\section{Introduction}
Core-collapse Supernovae (CCSNe) are one of the most energetic explosions in the universe. 
The release of gravitational energy reaches about $10^{53}{\rm~erg}$ and more than 90\% of that energy is thought to be released as neutrinos~\cite{1987PhLB..196..267S,Burrows:1988ba,1989ApJ...340..426L}. 
After the explosion supernovae leave behind compact objects, neutron stars or black holes. 
Due to extremely dense matter at the core of the supernova observation of star's interior during the explosion via electromagnetic signals is difficult, making it harder to understand the formation of those compact objects. 
However, since neutrinos rarely interact with matter they can carry information about the inner cores of the exploding star, such that 
their observation on Earth allows for the study of the explosion mechanism, the structure of neutron star, and the formation of black holes.
For this reason neutrino detectors all over the world are preparing to observe events from the next galactic supernova.

SK~\cite{Super-Kamiokande:2002weg} is a water Cherenkov detector located about 1000\,m under Mt. Ikeno in Gifu prefecture, Japan, and is continually monitoring for supernova neutrinos. 
SK is composed of a stainless steel tank filled with of 50.0\,kton ultrapure water. 
The tank is optically separated into two regions: an inner detector (ID) and an outer detector (OD).
The 32.5~kton ID is equipped with 11,129 inward-facing 20-inch photomultiplier tubes (PMTs) to observe Cherenkov rings formed by charged particles traversing its interior.
To identify incoming particles, such as cosmic ray muons, and to tag particles going out from the ID, the OD is instrumented with 1,885 outward-facting 8-inch PMTs 
coupled to wavelength-shifting plates.
Although typical SK analyses use an ID fiducial volume of 22.5\,kton to 27.2\,kton, the full volume of the ID is expected to be useable during supernova neutrino bursts~\cite{Super-Kamiokande:2022dsn}.

Based on the observations of neutrinos from SN~1987A~\cite{Hirata:1987hu,Bionta_1987,Alexeyev_1988}, humanity's only detection of a CCSN via neutrinos so far, 
thousands of neutrino events are expected to be detected with SK~\cite{Mori:2020ugr} for a galactic CCSN. 
For a closer SN, such as Betelgeuse, over tens of thousands of events are expected within a minute. 
Indeed, we consider the observation of Betelgeuse with the SK. Given that the Kamiokande experiment observed 11 events from SN~1987A which is located in the LMC at 51.2\,kpc, and the distance to Betelgeuse is $168^{+28}_{-15}$\,pc~\cite{2020ApJ...902...63J},
we anticipate that $11{\rm M} - 20{\rm M events}$ interactions will occur over 10 seconds within the ID full volume of SK.
This number of events is too many for the current SK data-acquisition (DAQ) system to process.
Accordingly, we have developed new DAQ modules specifically to enable recording of events from a very nearby supernova to address this problem. 
Two modules have been newly developed, one of which records only the PMT hit rate before any time and charge information is extracted from the standard DAQ and another that prescales the data passed to that DAQ when the event rate becomes large. 
The development and performance of these modules is discussed below.

\section{Current SK DAQ}

Prior to 2008 so-called Analog Timing Modules(ATMs) were used to digitize the analog PMT signals in SK following a hardware-based trigger decision. 
The ATM suffered dead time as a result of its slow digital conversion time as well as slow data processing and readout speeds. 
During the conversion and readout, only a maximum of subsequently triggered 2 events could be recorded.
To overcome these limitations, the SK front-end electronics were replaced with the QTC Based Electronics with Ethernet (QBEE) module, which 
continuously records all PMT hits in the detector, including dark noise and other spurious signals, and sends the data to a software-based 
trigger system. 
The QBEE-based DAQ system realizes dead-time-free data taking at lower energy thresholds than the ATM system.
Further, while the ATM is capable of processing a trigger rate up to 4\,kHz, roughly the rate expected from a CCSN at 5\,kpc, 
the QBEE can handle more than 20\,kHz. 

\begin{figure}
    \centering
    \includegraphics[width=18cm]{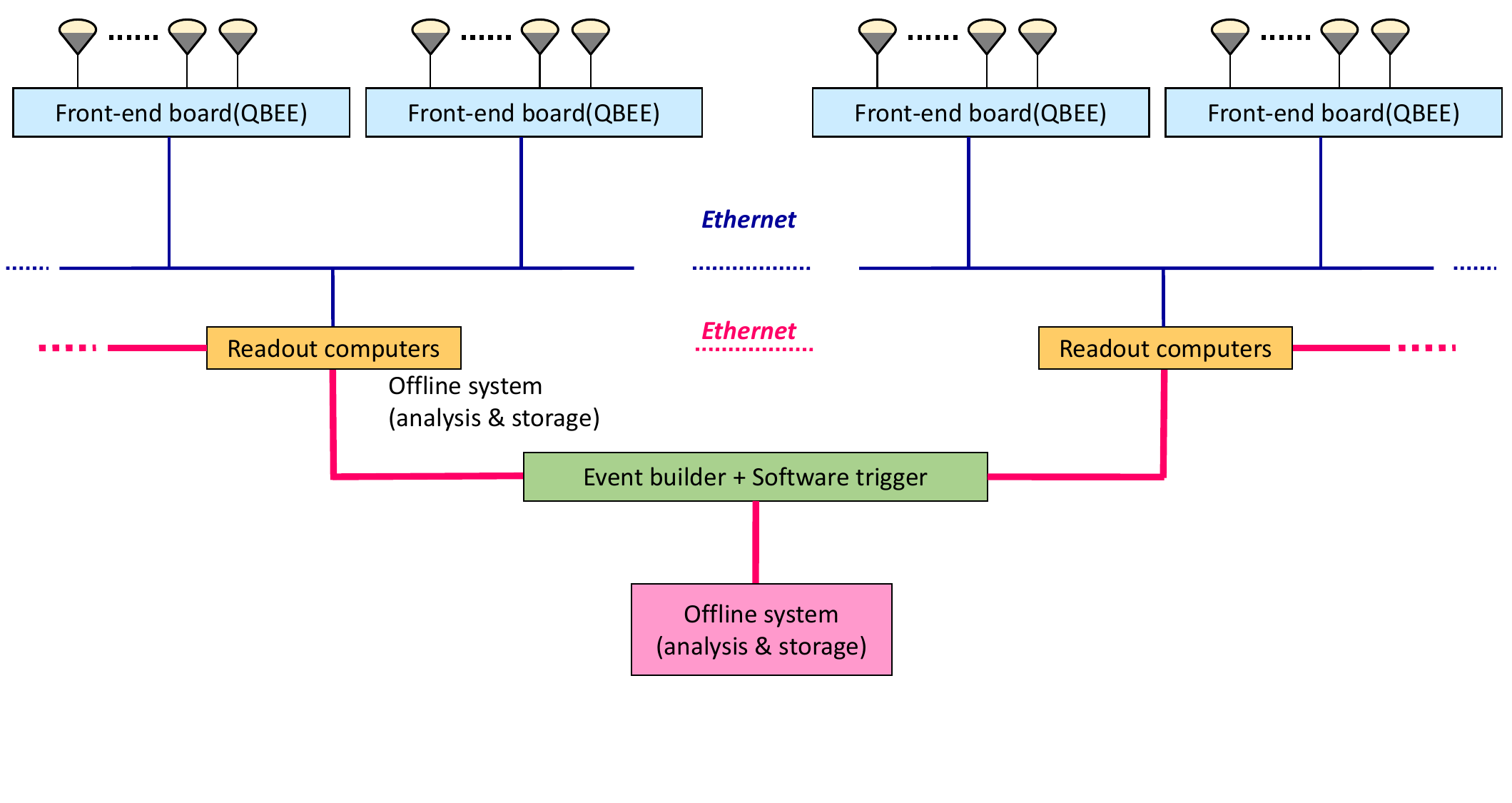}
    \caption{Schematic diagram of the flow of data in the SK DAQ system.}
    \label{fig:SK_data_flow}
\end{figure}

Figure~\ref{fig:SK_data_flow} shows the nominal data flow in SK since 2008. 
In total there are 550~QBEE boards each of which is connected to up to 24~PMTs and is stored in one of four huts at the top of the SK tank. 
These huts divide the detector into quadrants and also hold house HV power supplies for the PMTs as well as front-end readout computers. 
Each QBEE has eight charge-to-time converters (QTC)~\cite{Nishino:2009zu} and four time-to-digital converters (TDC (AMT (Atlas Muon TDC)))~\cite{Arai:2000eb} 
for digitization.
When relativistic charged particles traverse the SK tank the emitted Cherenkov light is converted to analog electrical pulses by the PMTs.
These analog pulses are converted to digital signals via the QBEE's onboard QTC before being processed with the TDC. 
After this step twelve readout computers collate the data from the entire detector into time series before 
passing them to eight computers dedicated to apply the software-based trigger to form events. 
Triggerd events are stored to disk after removing PMT hits occurring in multiple events.




All QBEEs are synchronized by a master-clock module (MCM) capable of distributing a 60~MHz and 60~kHz clock.
The 60~kHz clock signal is passed to an event-counting module (TRG32), which passes the generated event count (termed ``hardware counter'' elsewhere in this paper) 
back to the MCM for redistribution to the QBEEs. 
The 60 kHz clock is also used to reset the QBEE TDC thereby synchronizing the PMTs to the MCM clock. 
We note that a pedestal trigger is also generated using the 60\,kHz clock of the MCM.


In order to identify interesting regions in the data stream, the software trigger utilizes the number of PMT hits in a sliding 200\,ns time window to make trigger decisions. 
The criteria for various triggers are summarized in Table~\ref{tab:trigger}. 
Currently the DAQ system provides essentially dead-time free acquisition of all PMT hits above these thresholds. 
Supernova neutrinos are expected to generate primarily SLE, LE and HE triggered events.
For electrons with kinetic energies greater than 3.5~MeV the SLE trigger efficiency is 100\%, whereas 
above 5.4~MeV (8.5~MeV), the LE trigger efficiency is 50\% (100\%). 
Both efficiencies include the impact of the few kHz dark noise generated by the ID PMTs.

\begin{table}[htbp]
    \centering
    \begin{tabular}{ccc}
        \hline\hline
        Trigger Types & Criteria & Gate Width[$\rm \mu s$]\\
        \hline
        SLE &  $N_{200}>31$ & [-1.5,
        +1.0]\\
        LE & $N_{200} > 47$ & [-5,+35]\\
        HE & $N_{200} > 50$ & [-5,+35]\\
        SHE & $N_{200}>58$ &
        [-5,+35]\\
        OD & $N_{200} > 22$ in OD & [-5,+35]\\
        AFT & SHE + no OD & [+35,+535]\\
         \hline\hline
    \end{tabular}
    \caption{Summary of the SK event trigger menu, where $N_{200}$ is the number of PMT hits in a sliding 200\,ns time window. Here OD refers to the number of hit PMTs in the outer detector. These conditions are current as of 2023 and may vary according to the situation of the detector, for example, dark rate. \label{tab:trigger}}
\end{table}

\subsection{Limitations of the current system}
\begin{figure}
    \centering
    \includegraphics[width=12cm]{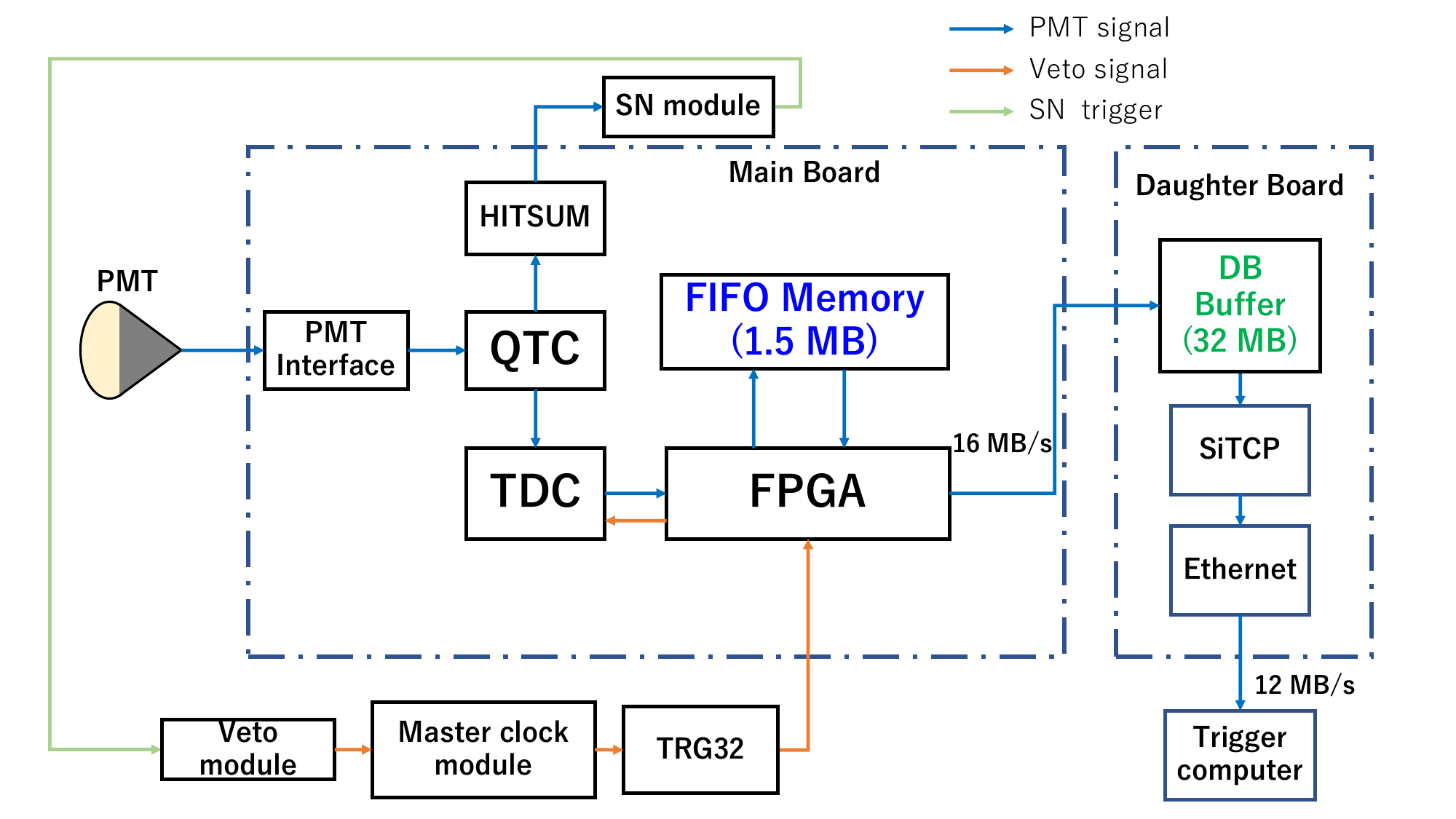}
    \caption{Simplified Block diagram of the QBEE.}
    \label{fig:QBEE_circuit}
\end{figure}

While the QBEE can handle higher event rates than the previous ATM module, for extremely large data rates, such as those expected from a nearby supernova, buffers on the board as well as downstream DAQ computers may overflow resulting in a loss data.
 Figure~\ref{fig:QBEE_circuit} shows a simplified block diagram of the QBEE board.
 Each QBEE is connected to PMTs via the PMT interface and has three circuits to measure the observed charge over separate ranges in order to keep linearity. 
 Analog pulses from the PMTs are converted by the QTC into rectangular pulses whose widths correspond to the total charge in the input signal.
 These rectangular pulses are then converted to digital signals by the TDC.
 A field programmable array (FPGA) located downstream of the TDC then converts the width between the rise and fall the a pulse into charge, divides the data into blocks of hardware counters and chooses a suitable charge integration range for the signal.
 Finally, the digitized data are transmitted to the readout computer through the daughter board (DB).

The QBEE board is equipped with three types of memory to store pulse information: TDC memory (L1 buffer), FIFO memory and a buffer on the daughter board.
The L1 buffer is used to store pulses within the TDC and it is the smallest of the three. 
Pulses being processed by the FPGA are stored in the FIFO memory and the onboard DB buffer stores data before being sent to the readout computers.
Under normal run conditions these buffer memories do not overflow, though it can happen in the L1 buffer 
following very high charge signals from energetic through going muons, that induce multiple rapid pulses in the PMTs. 
However, a supernova such as Betelguese, which is predicted to generate tens to thirty million events over 10 seconds, 
can cause overflow in these buffers.
When the buffers are full additional data is discarded, resulting in a loss of information available to reconstruct individual events.
\section{New modules for overflow protection}
%

 To overcome the difficulties described above, two new DAQ modules have been developed, the SN module~\cite{Orii:2016qvf} and the Veto module.
 Conceptually the Veto module serves to limit the amount of data that the QBEE processes during a very near supernova burst, taking as input 
 the PMT hit rate recorded by the SN module. 
 The SN module records the PMT hit rate without incurring any dead time. 
 Indeed, the hit rate serves as a proxy for the rate of energy deposition and even by itself is useful in the analysis of supernova data.
 These two modules make it possible for dead-time free data acquisition and for the prevention of uncontrolled data loss caused by QBEE memory overflows. 
 As explained below the SN module records only the number of hit PMTs continuously and is not subject to memory overflow issues. 
 When the Veto module detects a PMT hite rate that is too high for the QBEE, it issue a veto to stop QBEE processing, thereby prescaling
 the event rate it sees. 
 During this veto the QBEE does not store data. 
 However, for the low energies typical of neutrinos from supernovae, the number of hit PMTs is proportional to the neutrino energy so 
 the loss of QBEE data is compensated by the number of hit PMTs stored by the SN module.

 In order to extract the most information from a supernova burst in which the QBEE is periodically vetoed in this way, 
 we must rely on the PMT hitsum provided by the SN module and supplemental information from the QBEEs during periods where no veto is issued.
 Assuming that the distribution of neutrino energies does not change appreciably during the veto window, typically 17\,$\mu$s, the flux of neutrinos during that time can be extracted from the QBEE information recorded during the surrounding non-vetoed time periods. 
 In short, the QBEE's measurements of the number of PMT hits taken together with reconstructed event energies can be used to extract the flux given the PMT HITSUM provided by the SN module during the veto period.

 \subsection{SN module}
 The SN module is a dedicated system that records the total number of PMT hits (HITSUM) in the detector~\cite{Orii:2016qvf}. 
 This HITSUM is calculated by the QBEE, but importantly is done upstream of the QBEE memory storage as shown in Figure~\ref{fig:QBEE_circuit}.
For this reason the HITSUM calculation operates independently of the TDC and is not affected by memory overflows when the hit rate becomes large.
The SN module reads out the HITSUM at 60\,MHz from the circuit, sums up those hits at 60\,kHz and saves it together with the hardware counter associated to those hits.

\begin{figure}
      \centering
      \includegraphics[width=15cm]{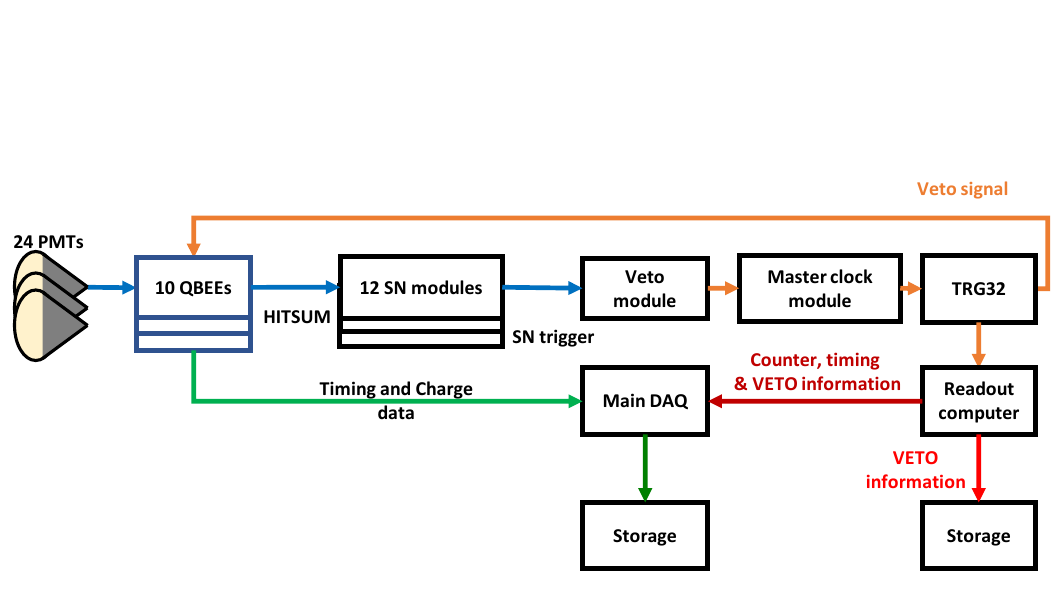}
      \caption{Relationship between the QBEE, SN module and Veto module as well as the flow of their data.}
      \label{fig:relation_new_modules}
  \end{figure}

Each of SK's electronics huts houses 12 SN modules and each board is attached to 10 QBEEs (Figure~\ref{fig:relation_new_modules}). The SN module issues a trigger when the HITSUM exceeds a programmable threshold. 
Currently one SN module will issue a trigger when there are more than 100 PMT hits in each of four consecutive hardware counters, roughly 68\,$\mu$s.
The Veto module, described later, uses the number of SN module boards issues triggers to determine when the QBEEs should be vetoed.

 \subsection{Veto module}

\begin{figure}
    \centering
    \includegraphics[width=12cm]{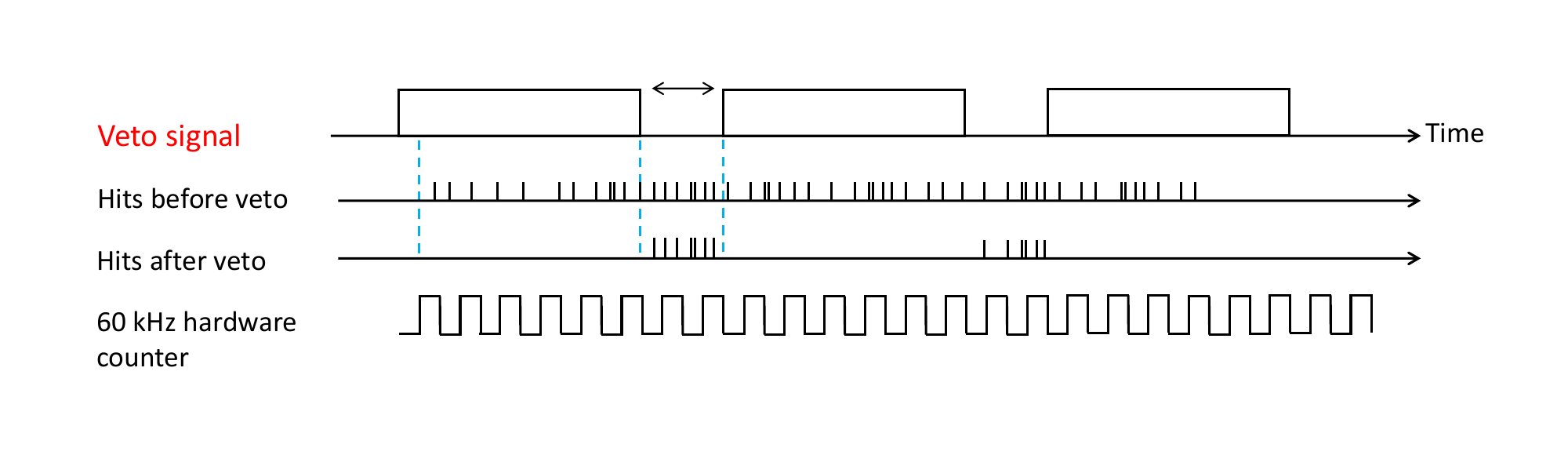}
    \caption{Time structure of the veto signal to implement a 1/4 prescale factor.}
    \label{fig:veto_time_structure}
\end{figure}

  The Veto module is designed to prescale the number of PMT hits to prevent the QBEEs from overflowing. The Veto module was officially put into operation in June 2021. \textcolor{black}{The Veto module was developed with a GNV-252 VME board, which has a single FPGA and four NIM ports that are controlled with firmware. However, the Veto module needs 14 NIM ports: one clock output, one veto signal output and twelve SN trigger inputs. Accordingly, we attached a GNV-260 expansion board which has 16 NIM ports. Both boards are commercially available.}
   Each Veto module is connected to 12 SN modules such that its veto decision is based on the activity of PMTs in each quadrant of the detector.
   For a nearby supernova the rapid succession of neutrino interactions in the detector is expected to produce a burst of PMT hit activity 
   that will trigger the SN modules.  
   Looking at those SN module triggers the Veto module vetos QBEE activity at and downstream of the TDC to prevent memory overflows. 
   Table~\ref{tab:veto_settings} lists the conditions under which the module issues a veto together with the corresponding prescale rates.
   \textcolor{black}{These conditions were determined with laser diode test data as described in the next section.  It is essential to prevent the QBEEs from overflowing because while the controlled data sampling by the Veto module can be compensated by information from the SN module, random data losses due to overflow can not be compensated. 
   For this reason, we adopt conservative trigger conditions to prevent overflow even during 60 million event bursts. 
   Such bursts produce two to three times as many as is expected from Betelgeuse becoming a supernova.}
  Figure~\ref{fig:veto_time_structure} shows the timing diagram issued veto signals. 
  Based on the activity in the detector the veto may be applied over several hardware counters, but when the veto is lifted there will be at least two hardware contours before another veto is possible. 
  Note that the Veto module runs with an internal clock and is not synchronized with the QBEEs. 
 
  When the event rate triggers the Veto module, a veto signal is first sent to the MCM (Figure~\ref{fig:QBEE_circuit}) whose primary function is to distribute clock signals and hardware counters to all QBEE boards. 
  Upon receipt of a signal from the Veto module though, the MCM then distributes the veto signal to all QBEEs using the next clock. 
  Veto signals are transmitted to the TDC on QBEE, stopping its function. 
  As a result, the TDC discards any received pulses during one 17\,$\mu$s cycle while the veto is applied.  
Note that the veto signals do not have any influence on HITSUM, which continues to process PMT hits since it is independent of the TDC. 
 Veto signals are recorded via the TRG32 in Figure~\ref{fig:relation_new_modules} independently of the main DAQ so that we exactly 
know when veto signals issue even if the QBEEs stop due to an overflow or due to the veto signals themselves. 
 


  \begin{table}
    \centering
    \scalebox{0.75}{
    \begin{tabular}{ccc}
    \hline\hline
         Threshold of SN trigger & Length of continuous excessive SN trigger & Prescaling rate\\
         \hline 
         8  & 4   &  1/4 \\
         10 & 100  & 1/5 \\
         10 & 1000 & 1/6 \\
         \hline
    \end{tabular}}
    \caption{Trigger conditions for issuing a veto with the Veto module.}
    \label{tab:veto_settings}
\end{table}

\section{Performance tests of the new modules}

The performance of the SN modules and Veto modules was evaluated using dedicated data designed to simulate the event rate expected during 
a nearby supernova.  
These tests confirmed that the data are sufficiently prescaled to prevent memory overflows and further that 
they have no influence on normal observation periods.

\subsection{Laser Diode testing}
A laser diode (LD) connected to a diffuser ball located near the center of the SK tank is used to approximately reproduce the expected event rate of a supernova. 
For the tests described in this subsection we modified the total event rate of the simulated supernovae, while keeping the timing structure constant, to 
probe the limitations of the main DAQ without the benefite of the new modules. 

\begin{figure}
    \centering
    \includegraphics[width=10cm]{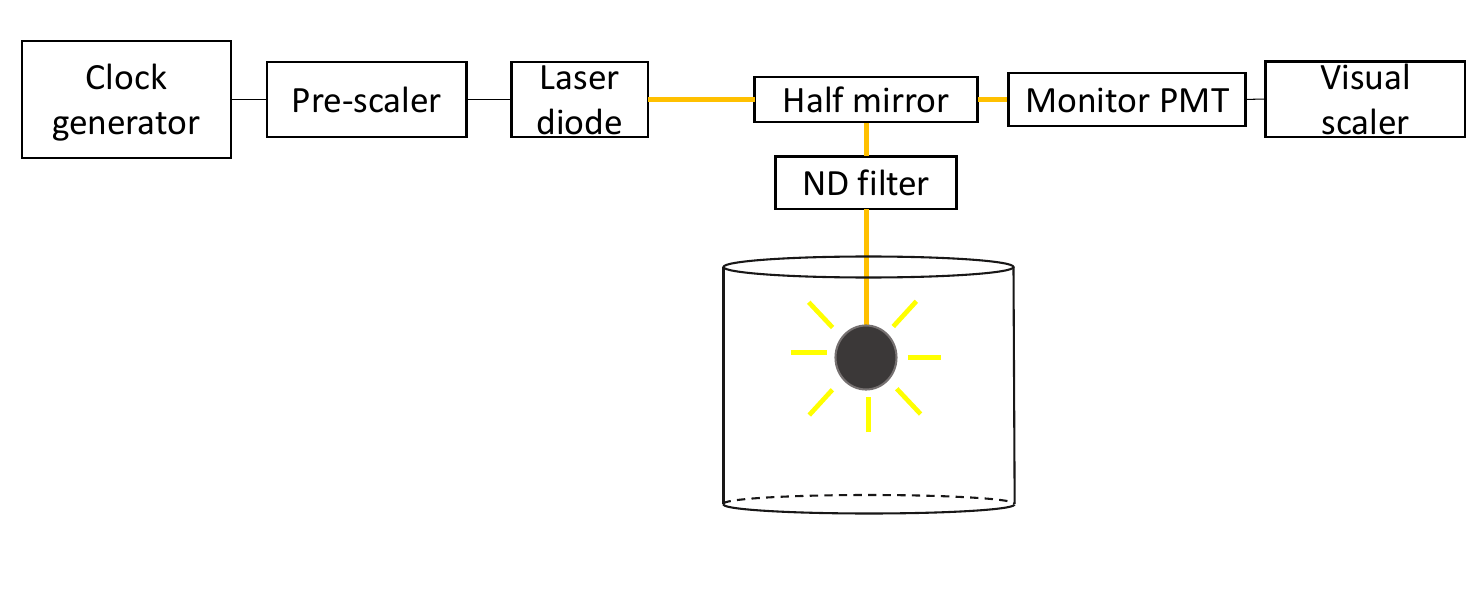}
    \caption{Setup of the LD test.}
    \label{fig:LD_test_circuit}
\end{figure}
\begin{figure}
    \centering
    \includegraphics[width=10cm]{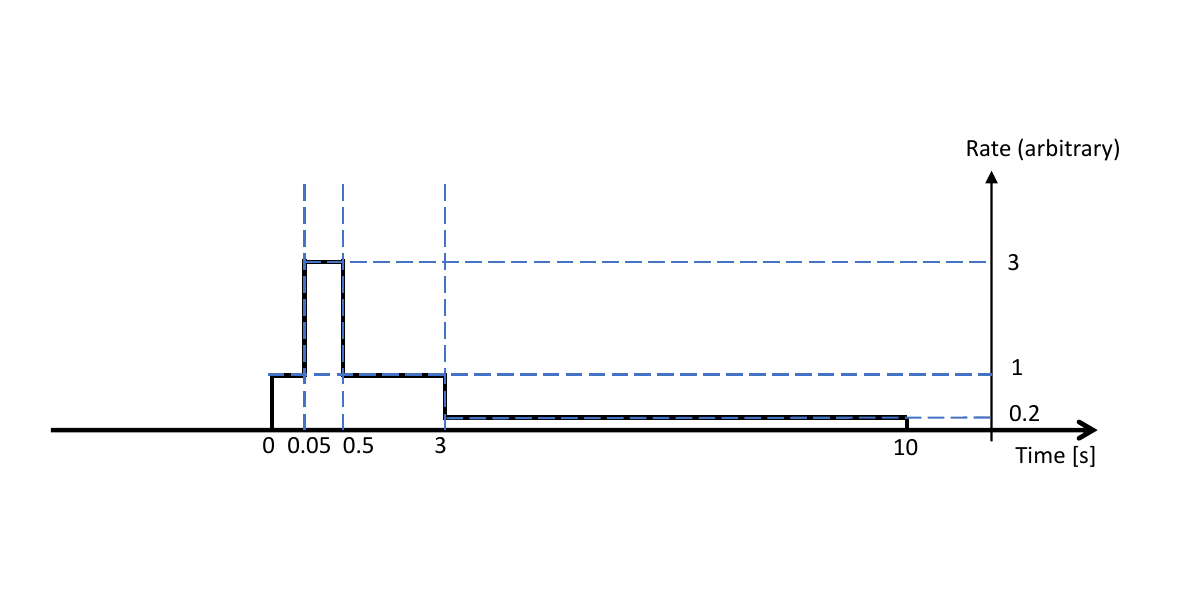}
    \caption
    {Time structure of the LD burst. Changing the total number of flashes, does not affect the horizontal time structure but instead increases the number of events in each section.}
    \label{fig:LD_time_structure}
\end{figure}

Figure~\ref{fig:LD_test_circuit} shows the setup of the LD test.
Light from the LD is isotropically emitted from the diffuser ball with the time structure shown in Figure~\ref{fig:LD_time_structure}.
A flash roughly generates 130 PMT hits, which correspond to 6\,MeV.
The time structure has been chosen to mimic that predicted by supernova models:
the first 0.5\,s corresponds to the sharp increase in the event rate expected from the neutronization burst, 
which decays rapidly over one second before the event rate gently slows for up to 10\,s during the neutron star cooling phase. 
Setting the LD system to produce 60 million events over the 10\,s burst is a sufficient test of the DAQ system as it exceeds 
the event rate expected from Betelgeuse (10--30 million events, depending upon the model).

\subsection{LD tests with and without the Veto module}

\begin{figure}
    \centering
    \includegraphics[width=10cm]{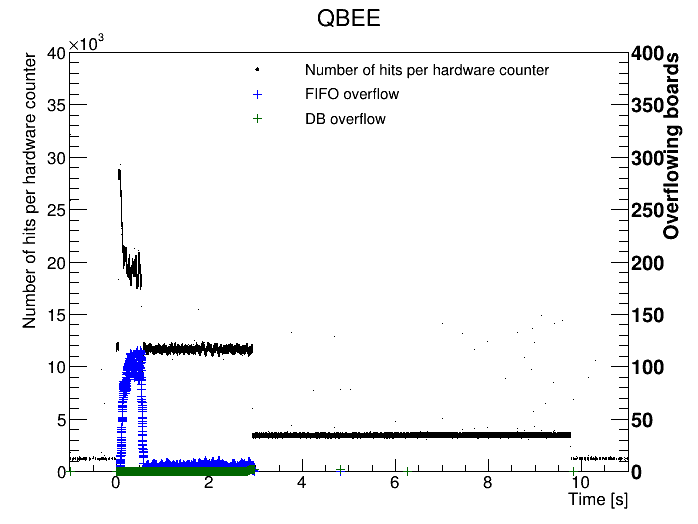}
    \caption{An LD test with 30 million burst events recorded with the QBEE. The black dots are the number of LD flashes per hardware counter ($\approx 17.1{\rm~\mu\,s}$).  Blue markers represent the number QBEE boards per hardware counter of which experience a FIFO buffer overflow. Green points show overflow for the daughterboard. The burst structure is disturbed by the overflow of the QBEE. \textcolor{black}{The time origin is the moment when the hits per hardware counter exceed 10,000. Hereafter we use the same time origin in LD tests.}}
     \label{fig:30M_burst_qbee_noveto}
\end{figure}

Without employing the Veto module and setting the total number of LD-generated events to 30 million results in the QBEE data shown in Figure~\ref{fig:30M_burst_qbee_noveto}.
The vertical axis shows the number of hit PMTs per hardware counter (17~$\mu s$) and the blue and green markers show the number of boards in which FIFO overflow and DB overflow occurs, respectively.
Comparing Figure~\ref{fig:30M_burst_qbee_noveto} and Figure~\ref{fig:LD_time_structure}, it is clear that the simulated burst structure is significantly affected
during periods of overflow. \textcolor{black}{The neutrino burst of the LD burst generates 16-56\,MB/s data and it make QBEE overflow within 0.03\,s to 1\,s. Then QBEEs actually start to overflow at 0.03\,s in Figure~\ref{fig:30M_burst_qbee_noveto}.}


\begin{figure}
    \centering
    \includegraphics[width=10cm]{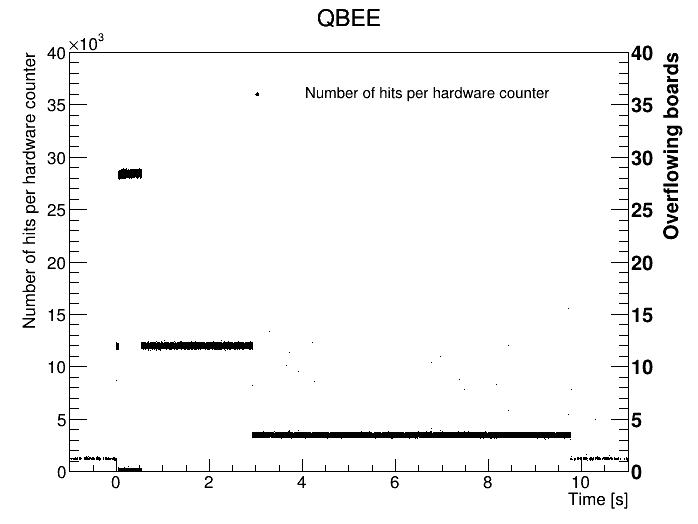}
    \caption{An LD burst test with 30 million events when the QBEE is connected to the Veto module. Here the burst shape is not disturbed. The cluster of zero hit from 0\,s to 0.5\,s is due to the Veto module. There is no QBEE overflow during the burst.
    }
    \label{fig:30M_burst_qbee_veto}
\end{figure}

\begin{figure}
    \centering
    \includegraphics[width=10cm]{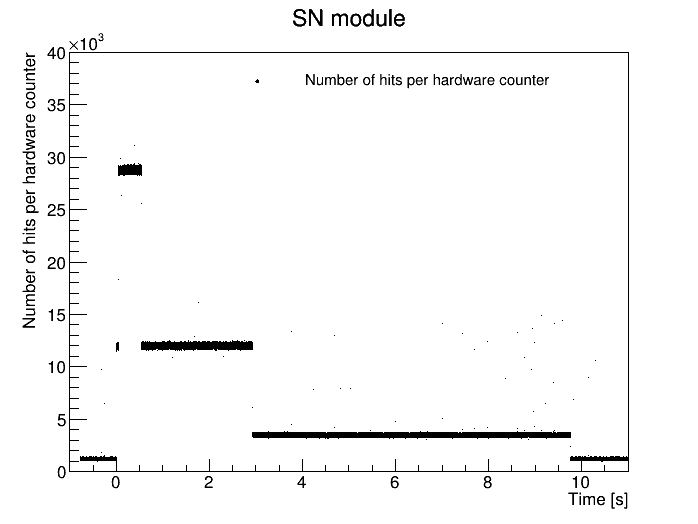}
    \caption{An LD burst test with 30 million events as recorded by the SN module. Note that this module can record the entirety of the LD burst.
    }
    \label{fig:30M_burst_snmodr_veto}
\end{figure}

Connecting the Veto module and repeating the test results in data shown in  Figure~\ref{fig:30M_burst_qbee_veto}.
Comparing again Figure~\ref{fig:30M_burst_qbee_noveto} and Figure~\ref{fig:30M_burst_qbee_veto}, we note that the Veto module completely prevents the QBEE from overflowing. 
Further, the time structure of the LD data is completely retained and  closely follows the original structure (Figure~\ref{fig:LD_time_structure}).  There is a cluster of activity near zero from \textcolor{black}{0\,s to 0.5\,s. This time period has the highest expected event rate and corresponds to QBEE hits being vetoed.}
Figure~\ref{fig:30M_burst_snmodr_veto} shows the same data set as recorded by the SN module, which is unaffected by the Veto module, as expected. 
Note that the baseline height in the SN and QBEE data are different because the latter records only triggered events, whereas the former continuously records all PMT hits.
\textcolor{black}{In Figures~\ref{fig:30M_burst_qbee_noveto}, ~\ref{fig:30M_burst_qbee_veto}, and ~\ref{fig:30M_burst_snmodr_veto}, there are sporadic times where the number of hits briefly exceeds the number in surrounding intervals. 
These isolated hits are due to cosmic ray muons which traverse the detector at a rate of about 3~Hz.}


\begin{figure}
    \centering
    \includegraphics[width=10cm]{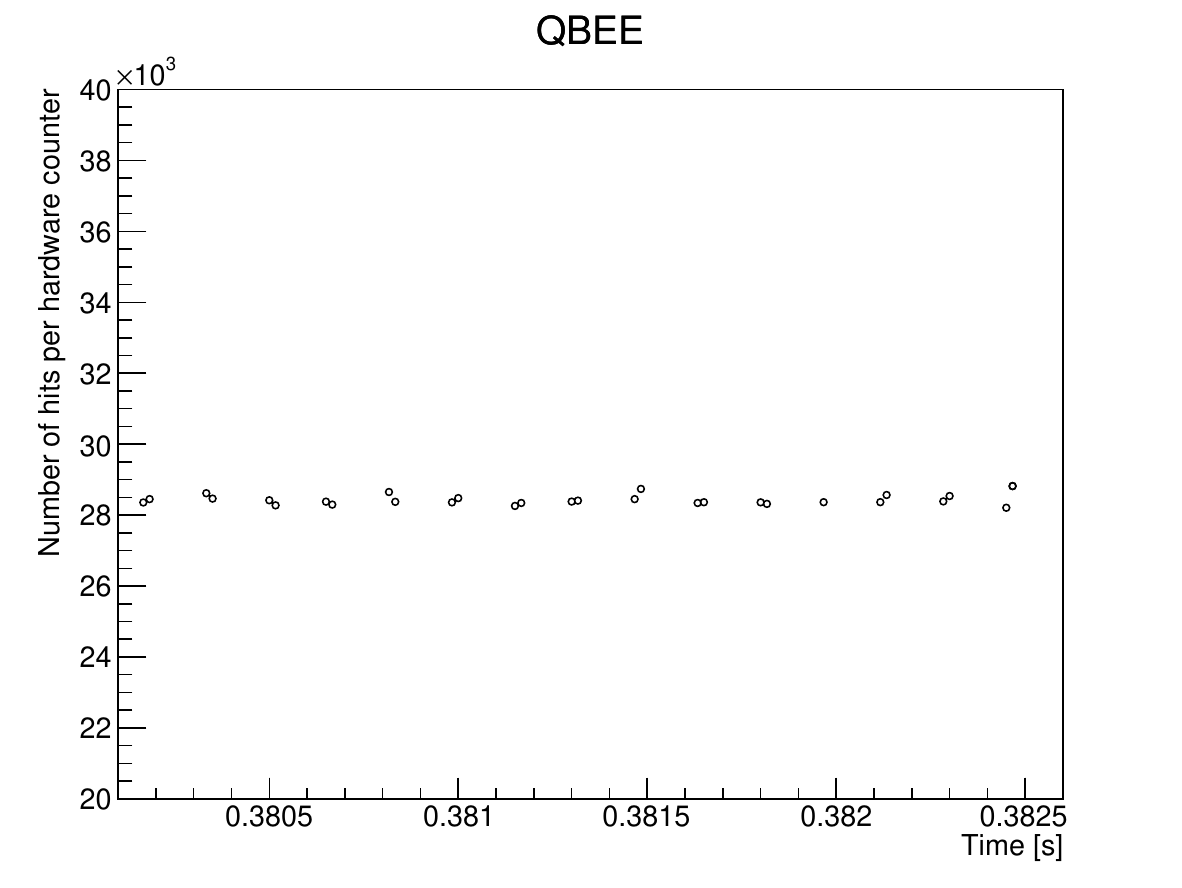}
    \caption{Zoom in of Figure~\ref{fig:30M_burst_qbee_veto} showing the structure of the veto signal. The veto length is 6 counters (103\,$\rm\mu s$) and the veto break length is 2 counters (34.1\,$\rm\mu s$). Paired dots are recorded PMT hits. There is an isolated dot at 0.382\,s due to a mismatch between the QBEE and Veto module clock signals.
    }
    \label{fig:30M_burst_qbee_veto_zoom}
\end{figure}

\subsection{Recorded hits from LD bursts}

Here we consider the results of four LD burst tests and compare their total number of hit PMTs. 
Defining the onset of a burst as the time when the number of hits in $17.1\,{\rm \mu s}$ exceeds 10,000 consistently for $170.7\,{\rm \mu s}$ and the baseline as the average of hits from 0.5\,s before the onset to the onset, we integrate the number of hits over 11\,s starting from 1\,s before the onset. \textcolor{black}{The average of the baseline is $1.215\times10^{3}$ hits and the standard deviation is $5.291\times10^{1}$\,hits.}
Table~\ref{tab:total_hits} summarizes the results.
The total number of hits recorded by the SN module with and without the veto module is almost the same, showing a difference of about 2\%. 
However, the QBEE recorded $2.266\times 10^{9}$ hits without the Veto module and $1.944\times 10^{9}$ hits with the Veto module. 
Indeed, the QBEE loses about 30\% of PMT hits due to buffer overflows for a burst with 30 million events, but 40\% of hits are vetoed Veto module. 
Though more events are seemingly lost due to the veto, we cannot use the remaining 70\% hits without Veto module for physics analyses as described in Section~\ref{sec:demonstration}. 
However, the hits remaining after vetoing the QBEE can still be used. 

\begin{table}
    \centering
    \begin{tabular}{ccc}
    \hline \hline
               &  QBEE  &  SN module \\
          \hline
    Veto module \textbf{off} & $3.340\times 10^{9}$ & $3.392\times 10^{9}$   \\
    Veto module \textbf{on} & $1.944\times 10^{9}$ & $3.318\times 10^{9}$ \\
         \hline \hline
    \end{tabular}
    \caption{Summary of the total number of hits recorded by each module during LD testing.}
    \label{tab:total_hits}
\end{table}

\section{Burst analysis demonstration}\label{sec:demonstration}
\begin{figure}
    \centering
    \includegraphics[width=10cm]{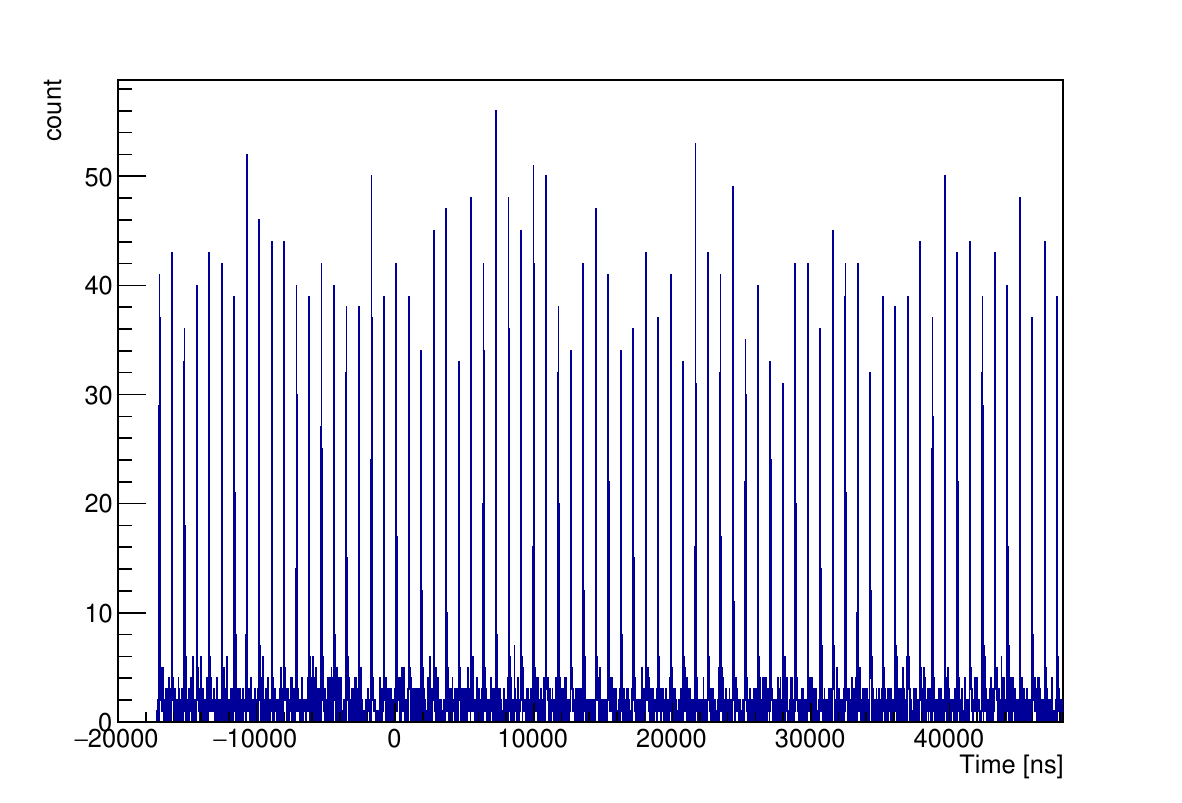}
    \caption{LD flashes in the tail of an event gate width. Each peak corresponds to an LD flash. The horizontal axis is time in nanoseconds and the vertical axis is the number of PMT hit counts per 16~ns.}
    \label{fig:hit_timing_one_ev}
\end{figure}

During the LD tests described above, each flash of the LD is meant to correspond to a neutrino event from a nearby supernova. 
In this section we analyze LD test data to reconstruct the simulated supernova timing structure. 
Figure~\ref{fig:hit_timing_one_ev} shows PMT hits from the tail of an LD test with a total of 30 million flashes. 
During the tail phase the flash rate is about 1.1~MHz, resulting in roughly 90 flashes during the 60~$\mu$s gate width 
assigned to events in SK. 
Typically only one event per such gate width is expected during normal operations and pile-up of this sort is not considered 
in ordinary SK analyses.

\begin{figure}
    \centering
    \includegraphics[width=10cm]{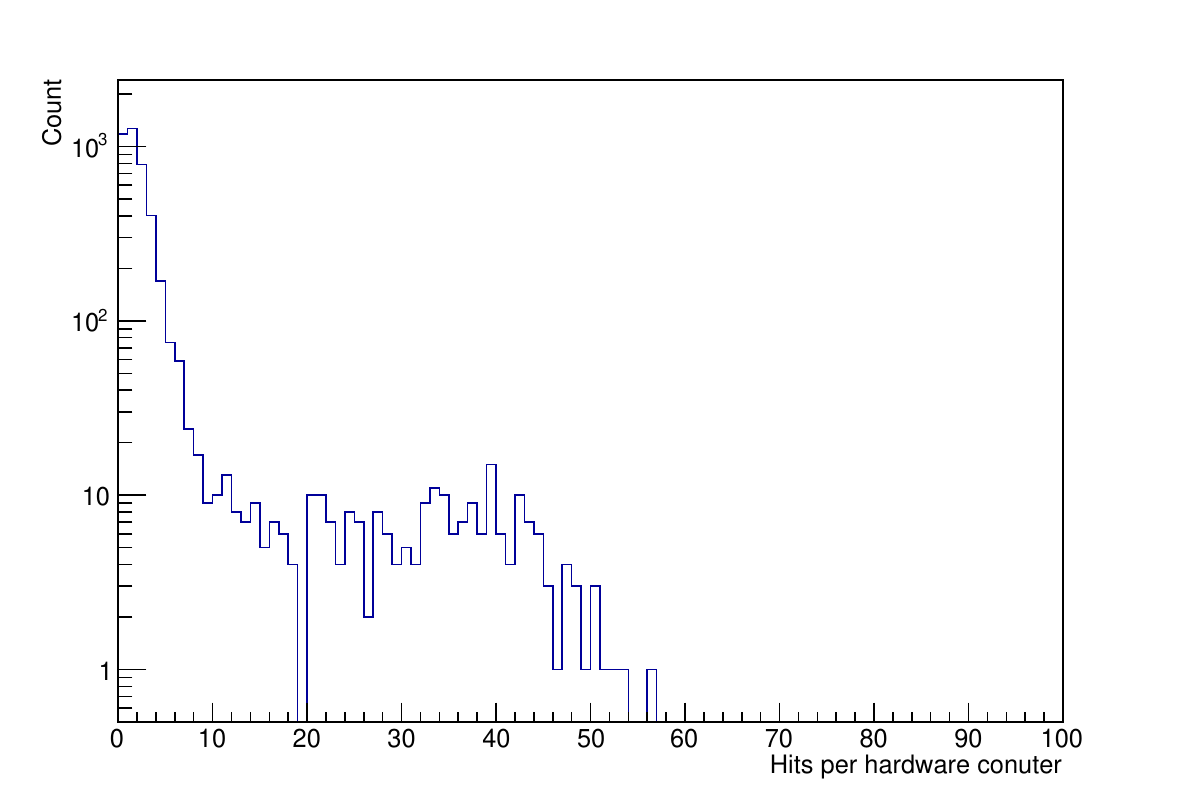}
    \caption{Histogram of LD test hit. LD flashes and dark noise are separated at about 20. There is a dip at 19 due to statistical uncertainty.}
    \label{fig:hit_one_ev}
\end{figure}

\begin{figure}
    \centering
    \includegraphics[width=10cm]{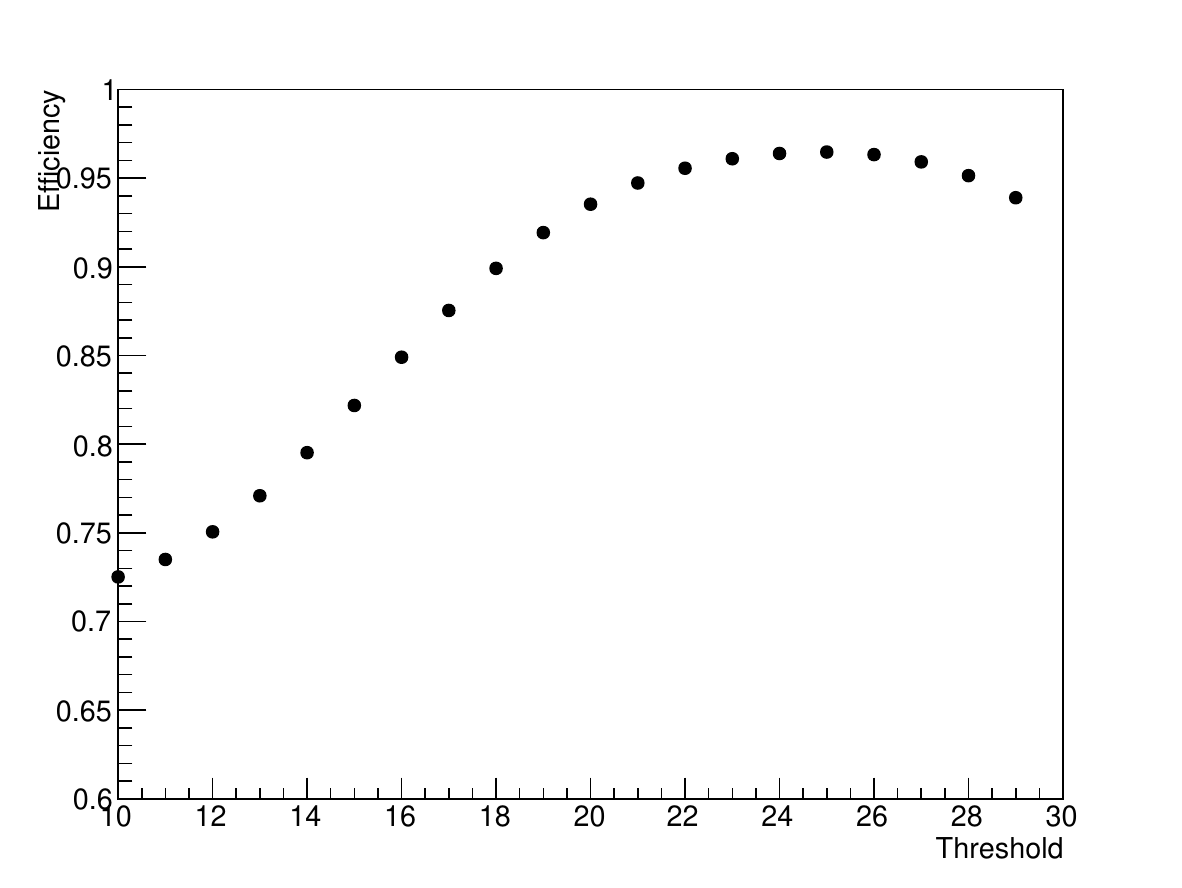}
    \caption{Efficiency of identification of LD flashes. The efficiency is the number of total flashes exceeded the threshold divided by 29.9M flashes.}
    \label{fig:eff_nth}
\end{figure}

\begin{figure}
    \centering
  \includegraphics[width=10cm]{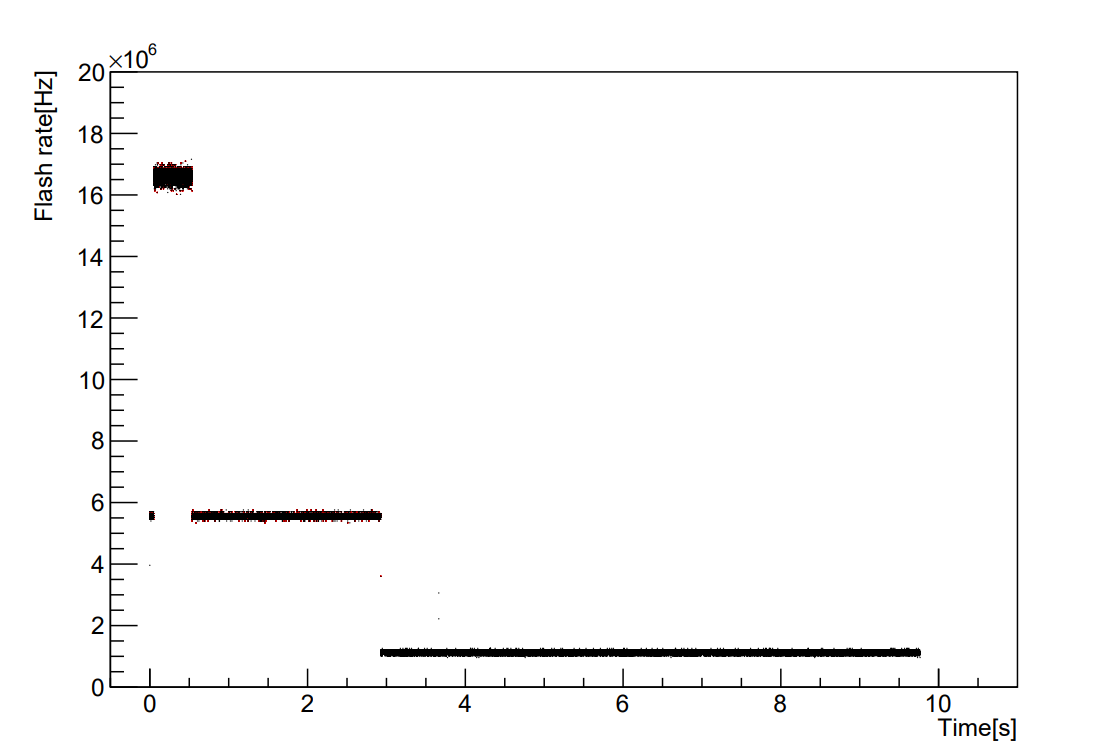}
    \caption{Flash rate of the LD test with 30 million events. The horizontal axis is time in seconds and the time origin is the same as in Figure~\ref{fig:30M_burst_snmodr_veto}. The vertical axis is the number of LD flashes per second.}
    \label{fig:flashes_rate}
\end{figure}

\begin{figure}
    \centering
    \includegraphics[width=10cm]{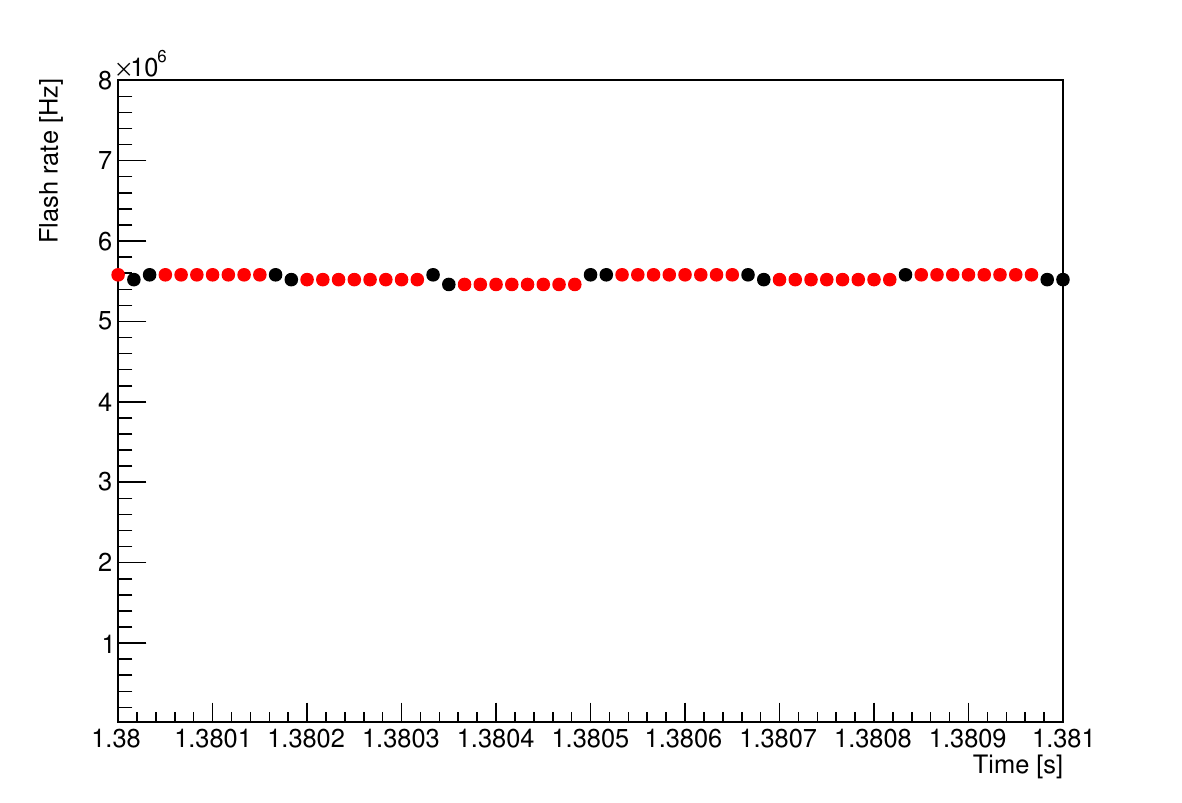}
    \caption{Zoom in of the area around 1.38~s of Figure~\ref{fig:flashes_rate}. The black points are data from the QBEE and red points are extrapolated points based on  the preceeding black points.}
    \label{fig:flashes_rate_zoom}
\end{figure}

Here LD burst data are separated by the hardware counter and the number of flashes during each counter is calculated by 
counting the number of PMT hits within a 68.25$\,\rm \mu s$ gate.
\textcolor{black}{
Figure~\ref{fig:hit_one_ev} shows the hit distribution of the LD test data. 
The population at low values is predominantly dark noise, while higher values are from LD flashes. 
Figure~\ref{fig:eff_nth} shows the efficiency for selecting an LD flash as a function of the number PMT hits used as a cut threshold. 
The efficiency is maximal for LD tests with around 25 hits, so in this analysis events with larger hit clusters are considered to be LD flashes.
When this threshold is less than 21 hits, an LD flash cannot be separated random dark noise or from a pile-up hits from a previous flash.
In this case, multiple LD flashes are likely to be regarded as a single flash which results in a decreased efficiency. 
Similarly, when the threshold is higher than 29 true LD flashes start to be rejected.
}
The number of flashes expected during vetoed hardware counters is calculated by interpolating 
based on the reconstructed flash rate in the previous hardware counter.
Figure~\ref{fig:flashes_rate} shows the reconstructed LD burst and 
Figure~\ref{fig:flashes_rate_zoom} shows the same data zoomed in around the 1.38\,s period.
The black dots are the real data taken by the QBEE and red dots show interpolated flashes from the previous data.

Summing the reconstructed flashes in Figure~\ref{fig:flashes_rate} from 0 to 10\,s results in a total of 28.8M flashes.
This value is close to the 30 million flashes expected.
The difference is due to pile-up of the flashes themselves during the highest flash rate of 0.5\,s. 
Such pile-up events are reconstructed as a signal flash with the procedure above.
Additionally, the expected number of flashes is based on the number of pulses sent to the LD.
It is possible that the LD did not flash during the highest event rates, so the number of reconstructed 
flashes may be closer to the truth than the number 30 million is. 
For example, the number of clocks to the LD is 29.9M. On the other hand, the number of counts reconded by the monitor PMT in Figure~\ref{fig:LD_test_circuit} is 28.6M events.

While analyses like the above are simple, we can nonetheless obtain physics information from them. 
Knowing the relationship between the number of hit PMTs and the deposited energy from calibrations, the neutrino flux during the SN burst can be estimated without detailed time and charge information from the PMTs. 
When the QBEEs are not vetoed, they provided detailed event information as seen in Figure~\ref{fig:hit_timing_one_ev}, such that 
if we assume the time evolution of the SN does not change much during the veto periods the neutrino flux therein can also be estimated 
by extrapolation from the surrounding QBEE data. 
Since the most rapid supernovae flux evolution is much longer than the typical veto length, milliseconds ~\cite{2014ApJ...786...83T,Nakazato_2013,Mori:2020ugr}) compared to tens of microseconds, this extrapolation should be robust. 

\section{Long-term Testing}

\begin{figure}
    \centering
    \begin{minipage}[t]{0.48\textwidth}
\centering
\includegraphics[width=8cm]{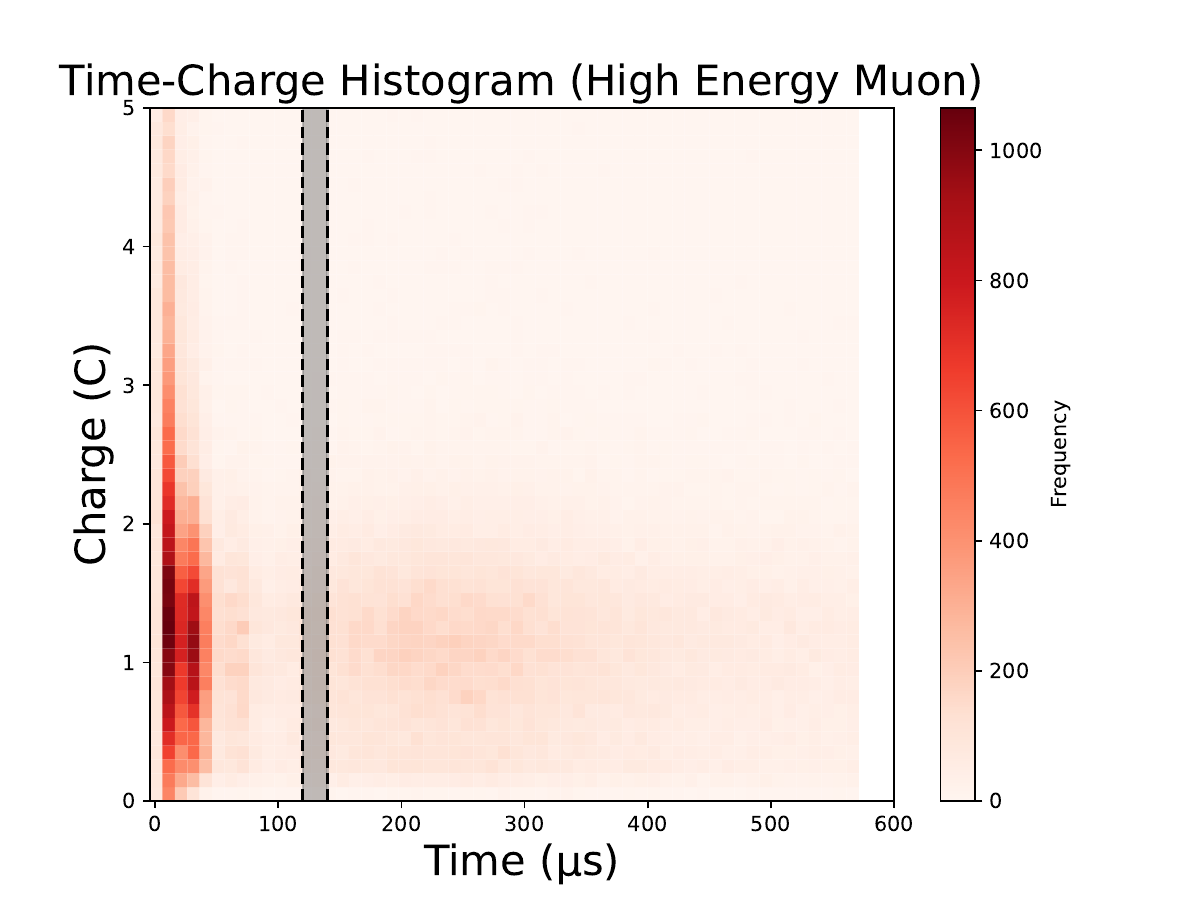}
\end{minipage}
\begin{minipage}[t]{0.48\textwidth}
\centering
\includegraphics[width=8cm]{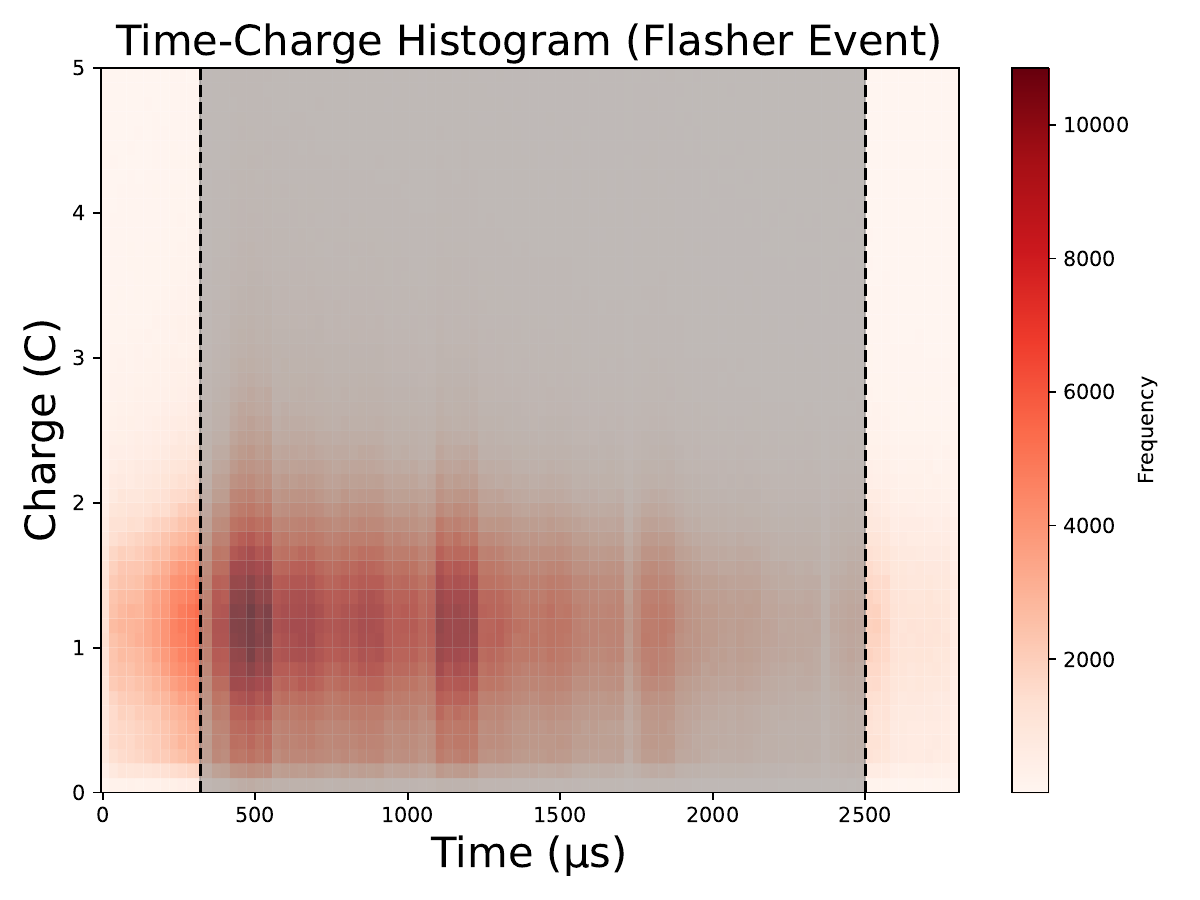}
\end{minipage}
\caption{Time-Charge histogram of hit PMTs from a high energy muon event (left) and from a flashing PMT (right). 
Gray bands indicate periods where the vetos were issued.
The muon event has a delayed veto signal of approximately 200\,$\rm\mu s$ due to inherent DAQ module latency. 
}
\label{fig:TQ_map}
\end{figure}

As the Veto module is a critical component of SK's ability to observe even the highest event rate supernovae, it must operate continuously and without interferring with normal DAQ operations. 
For this reason we conducted a year-long test of the module without allowing it to veto the normal DAQ even when the module was triggered.
Data from the long-term Veto module test was collected from July 2021 to June 2022.
Using recorded information of the veto signal (``storage'' in Figure~\ref{fig:relation_new_modules}), 
times were matched with events from the standard DAQ to identify what types of events triggered the Veto module. 
The event type is determined using the SK event display and Time-Charge histograms like that in Figure~\ref{fig:TQ_map}.


We observed two types of events which trigger the Veto module: highly energetic muons and PMT-generated flashers. 
Cosmic ray muons enter the tank at approximately 2\,Hz and some deposit enough energy to both saturate the PMTs and generate secondary particles through radiative losses. 
This may cause both a large number of hit PMTs in a short period of time as well as a significant amount of delayed hits from afterpulsing, which can result in satisfying the conditions to trigger the Veto module. 
PMT-generated flashers, on the other hand, are a phenomenon that occurs when a PMT emits a strong flash due to a discharge inside the PMT or the voltage divider.  
Repeated flashing of this sort, together with events produced by the reflections of those flashes can also mimic the event rate expected for a nearby supernova.

\begin{figure}
    \centering
    \includegraphics[width=13cm]{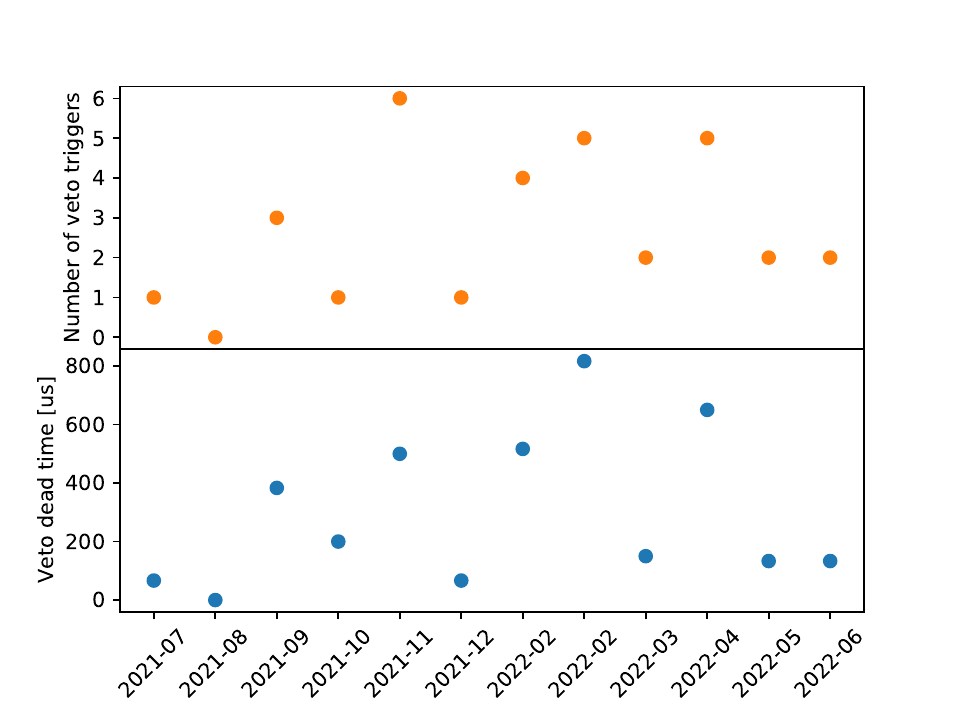}
    \caption{Triggering frequency of the Veto module together with the dead-time incurred during physics runs due to those vetos taken from July 2021 to June 2022. The top panel is the number of triggered Veto modules and the bottom panel is the dead-time. 
    }
    \label{fig:veto_time_distribution}
\end{figure}

Figure~\ref{fig:veto_time_distribution} shows histgrams of the number of triggered Veto module and the veto dead-time from July 2021 to June 2022. 
We found that the Veto module triggered 33 times during the long-term test, resulting in 3616\,$\rm \mu s$ of veto dead-time.
Looking at the Charge-Time profiles of those events, 30 are classified as highly energetic muons and one event is identified as a PMT-generated flasher.
The left panel of Figure~\ref{fig:TQ_map} shows the profile of highly energetic muon whose event length is 50\,${\rm\mu s}$,
The right panel shows the PMT-generated flasher which continues for 2500\,$\rm\mu s$ and produces three bands of activity with a period of 600\,$\rm\mu s$. 
These periodic bands are characteristic of PMT-generated flashers. 



Based on this test we expect roughly three vetos per month with an average deadtime per event of 100\,$\rm\mu s$.
Over the course of a full year approximately three miliseconds of data will be lost to vetos, which is negligible compared to the detector livetime. 




\section{Analysis of veto dead-time using Monte Carlo simulations}

Though we proved in previous sections that Veto module can prevent the QBEE from overflowing, allowing SK to take quality data during 
high event rate periods, the Veto module itself also creates some dead time via its vetoes. 
We have developed a new Monte Carlo (MC) simulation to investigate how much data well be vetoed by the module in response to supernovae simulated at 
various distances from the Earth. 

\subsection{Method}

Since 2020 SK has been loaded with gadolinium sulphate (0.01\% in 2020 and 0.03\% in 2022) 
and operated as SK-Gd~\cite{Super-Kamiokande:2021the}.
The purpose of the Gd loading is to enhance neutron detection and to thereby help separate inverse beta decay (IBD) events,
$\bar{\nu}_{\rm e} + p \rightarrow e^{+} + n$, 
from other reactions.
In water Cherenkov detectors IBD is the dominant interaction mode expected from supernovae, accounting for 90\% of the observed neutrinos. 
When neutrons are captured on gadolinium nuclei, a cascade of multiple gamma rays with 8~MeV of energy in total are emitted. 
While this makes the neutron capture easy to identify it creates additional, delayed PMT hit activity that is likely to affect the 
Veto module's operation. 

In our simulation studies we therefore consider both a pure water target and a gadolinium-loaded-water target.
The process includes inputting the supernova model, supernova distance and coordinates, and MSW neutrino oscillation assumptions~\cite{dighe2000identifying} into the SK event vector generator for supernova simulations called SKSNSim (Super-Kamiokande Supernova Simulation).
The software generates neutrino events according to input cross sections for IBD, charged-current interactions on oxygen, and neutral current interactions on oxygen 
convolved with the flux expected from the chosen supernova model. 
Output from SKSNSim is passed to the Geant4-based SK detector simulation, SKG4 (Super-Kamiokande Geant4)~\cite{harada2020geant4}, 
which simulates the passage of the outgoing particles through the detector. 
It simulates the passage of Cherenkov radiation through water (or the Gd solution~\cite{10.1093/ptep/ptz002, 10.1093/ptep/ptaa015}), including photon scattering, absorption, and reflection, and simulates the PMT and electronic responses~\cite{Nishino:2007ccp}. 
All analyses using Gd-loaded water in this study assume a 
 0.03\% Gd concentration~\cite{Super-Kamiokande:2021the}.
Afterwards realistic dark noise taken from random trigger data from SK is added to the output of the detector simulation. 
These simulated events are then analyzed to reconstruct the number of SN triggers and to determine when and for how long the Veto module would issue vetos.



\subsection{Supernova Models}

Due to the expensive computational resources required for supernova burst simulations, supernova models typically only simulate up to about 1 second after the bounce in multi-dimensional calculations. 
To overcome computational resource constraints, simplified assumptions and approximations (e.g., one-dimensional methods in a spherically symmetric geometry) are necessary for simulating neutrino emissions over long times. 
This study chooses two long-term one-dimensional supernova models: the Nakazato model~\cite{Nakazato_2013} and the Mori model~\cite{Mori:2020ugr}. 
Both models have flux predictions reaching up to 20\,s after the core bounce. 
An important distinction is that the luminosity of neutrinos predicted by the Nakazato model is higher than that of the Mori model.
Consequently, we expect the veto dead-time generated by the Nakazato model to be longer than that of the Mori model. 
When simulating supernovae from these models, we select distances ranging from 100\,pc to 1\,kpc at intervals of 100\,pc.


\subsection{Results and Discussion}

\begin{figure}
    \centering
\centering
\includegraphics[width=8cm]{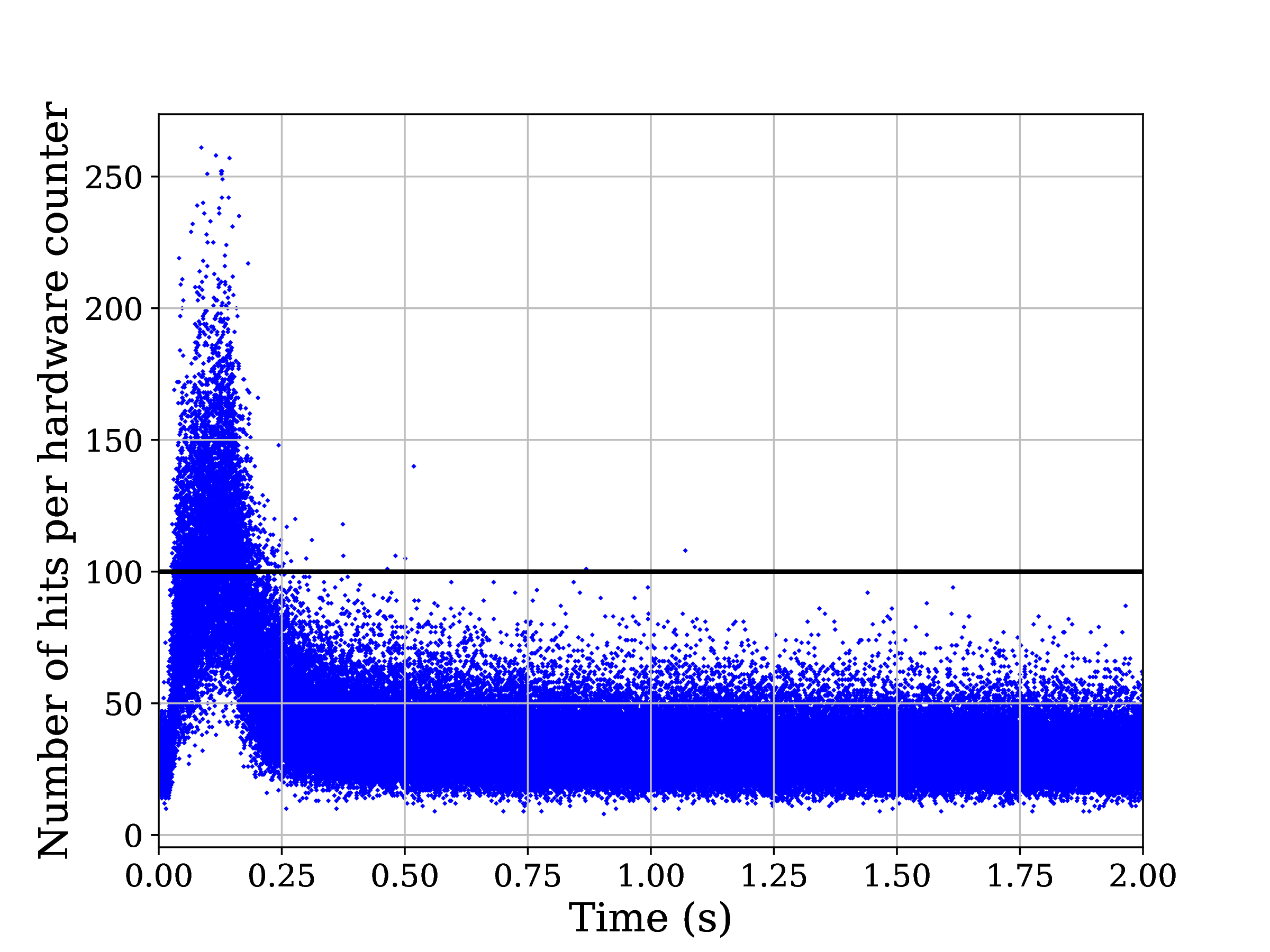}
    \caption{ Hits per a hardware counter as measured by the SN module for a supernova burst at a distance of 800\,pc from the Earth assuming the Nakazato model. The horizontal axis is the time measured from the first hit. The peak area meets the trigger condition of the SN module.}
    \label{fig:sncheck}
\end{figure}

\begin{figure}
    \centering
\centering
\includegraphics[width=8cm]{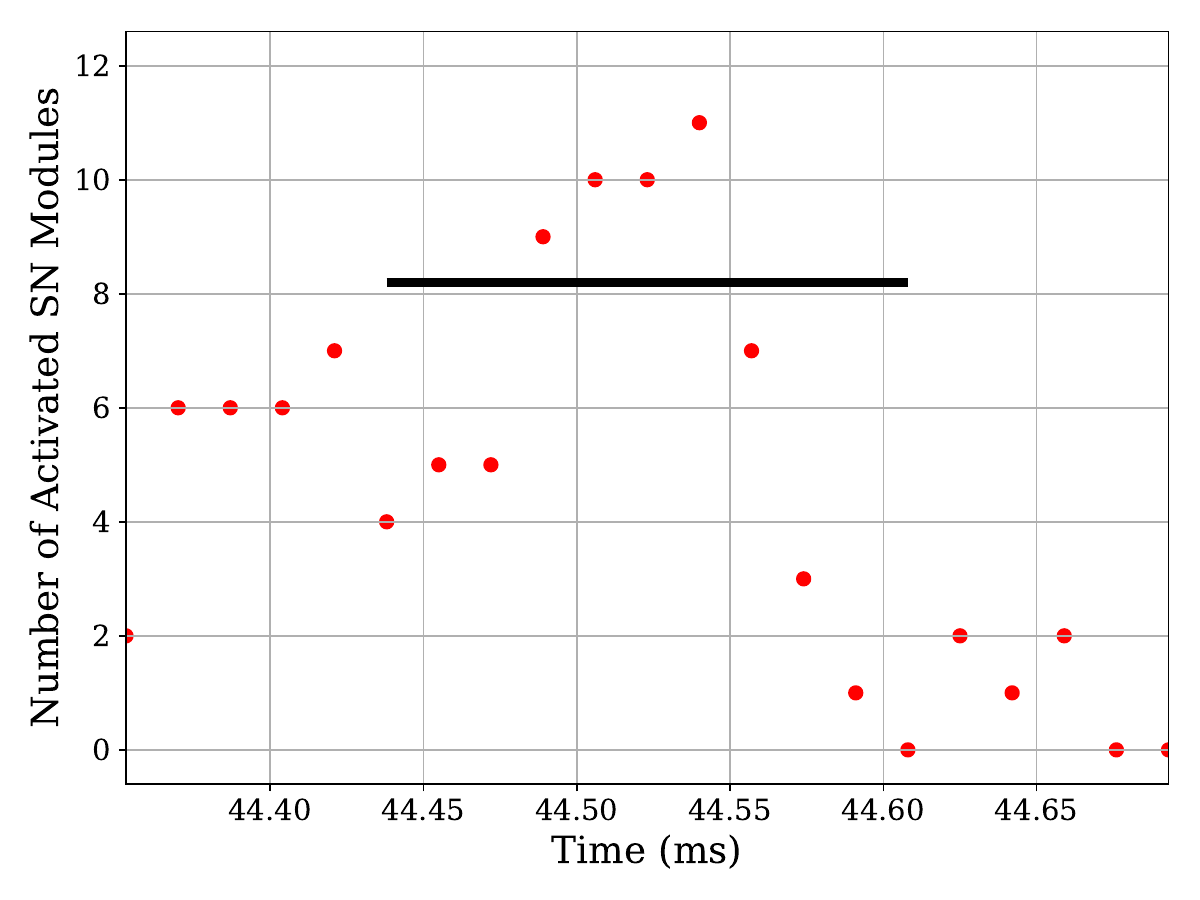}
    \caption{ SN triggers for the 800\,pc supernova of the Nakazato model. The Veto module conditions are listed in Table~\ref{tab:veto_settings}. The horizantal axis is  an expansion of the 0.4435\,s to 0.447\,s period in Figure~\ref{fig:sncheck}.}
    \label{fig:vetocheck}
\end{figure}

Figure~\ref{fig:sncheck} shows the number of PMT hits per 17.1\,$\rm \mu s$ recorded through one SN module for a simulated supernova located at 800\,pc from the earth as described by the Nakazato model. 
The distribution peaks between 0.03\,s and 0.2\,s hits during which the SN trigger condition of more than 100 hits continuing for 68.3\,$\rm \mu s$ or more is satisfied.
Figure~\ref{fig:vetocheck} then shows the issued SN triggers for the same supernova for the region from 0.443\,s to 0.447\,s. 
In this window and there are SN triggers which meet the condition of Veto module around 0.445\,s shown in Table~\ref{tab:veto_settings}, and during which 
the QBEEs would be vetoed by the Veto module. 

\begin{figure}
    \centering

    \centering
    \begin{minipage}[t]{0.48\textwidth}
\centering
\includegraphics[width=8cm]{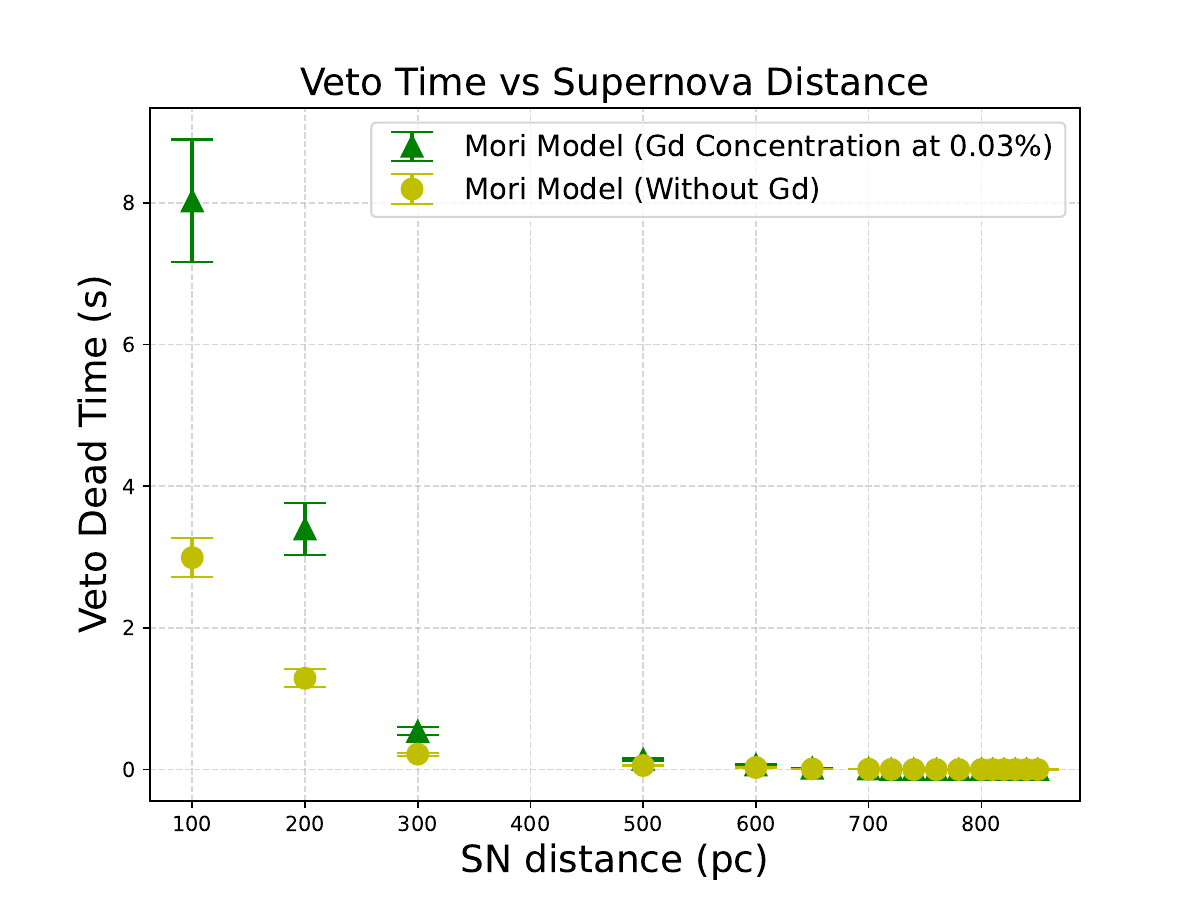}
\end{minipage}
\begin{minipage}[t]{0.48\textwidth}
\centering
\includegraphics[width=8cm]{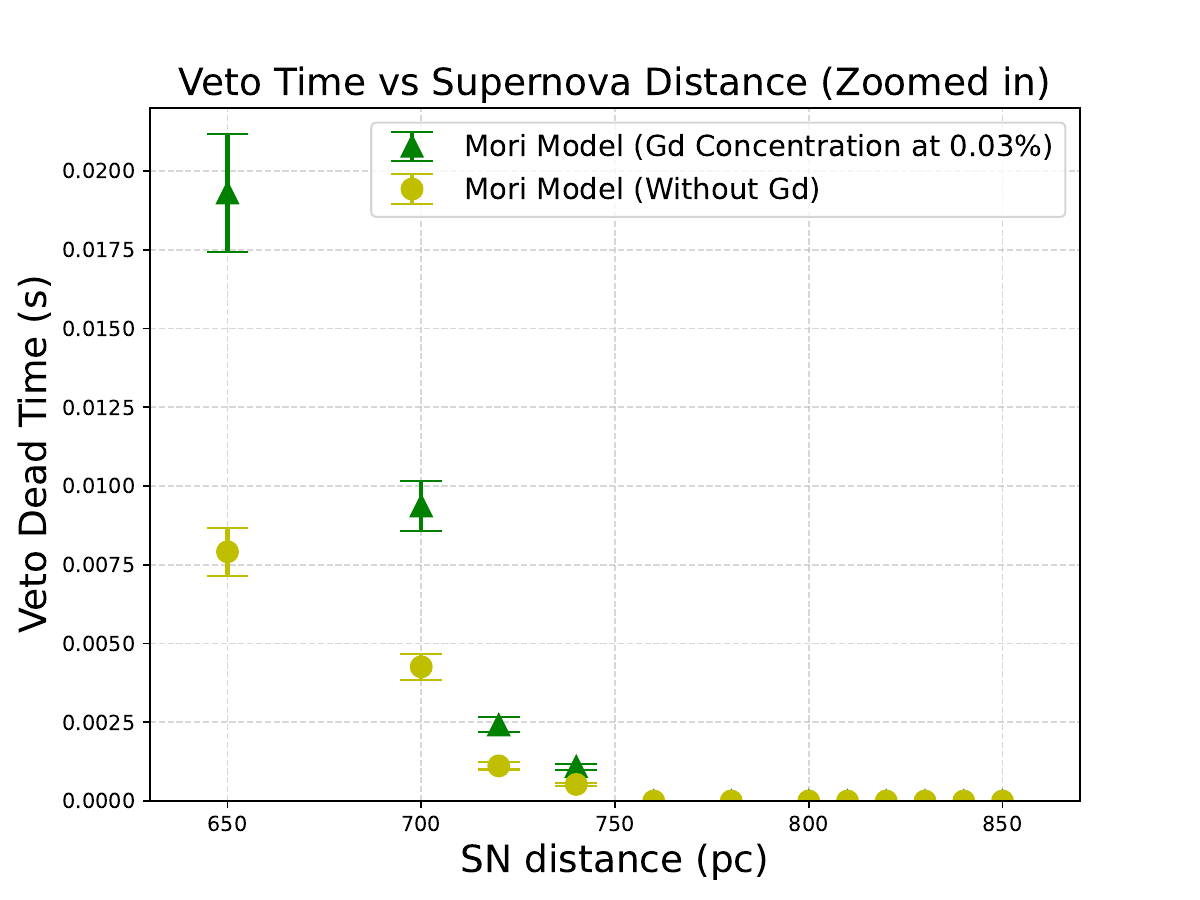}
\end{minipage}
    \caption{ Dead time incurred by vetoes issued by the Veto module as a function of simulated supernova distance. 
    Results from the entire survey of distances, from 100\,pc to 1000\,pc, are shown in the left panels, and enlarged views  
    of distances between 550\,pc and 900\,pc are shown in the right panels. 
    The error bars show $1\sigma$ statistical uncertainties. 
    It should be noted that 170\,pc is the distance between Betelgeuse and Earth.}
    \label{fig:MC_results}
\end{figure}

In this way, the amount of dead-time incurred by supernovae as a function of distance is calculated and summarized in Figure~\ref{fig:MC_results}.
The simulation results show that supernovae begin to trigger the Veto module at distances between 700\,pc and 850\,pc. 
For SK-Gd the module is triggered by more distant supernovae than pure water due to the gamma rays emitted when neutrons are caputured by Gd.
The Gd doping increases the total veto dead-time by approximately a factor of 1.6 regardless of the supernova's distance. 
Comparing the Nakazato and Mori models, we find that the distance at which the Nakazato model triggers Veto module is closer than that for the Mori model, 
as expected by the higer luminosity in the former. 
Table~\ref{tab:Sim_settings} summarizes the distances at which veto signals start being issued. 
With a 0.011\% concentration of Gd, this distance is closer by 20-30\,pc and the start of the Nakazato model is closer by 100\,pc than that of the Mori model.



\begin{table}[h]
    \centering
    \begin{tabular}{@{}lcS[table-format=3]@{}}
        \toprule
        Supernova models & Waters & {Start distances (pc)} \\
        \hline
        Nakazato  & 0.03\% Gd & {850 -- 900} \\
        Nakazato  & pure water & {820 -- 830} \\
        \addlinespace
        \midrule
        \addlinespace
        Mori  & 0.03\% Gd & {730 -- 740} \\
        Mori  & pure water & {710 -- 720} \\
        \bottomrule
    \end{tabular}
    \caption{Distances at which the Veto module starts being triggered for the Nazakato and Mori models assuming the pure water SK and SK-Gd configurations.}
    \label{tab:Sim_settings}
\end{table}

\section{Summary}
This paper describes new DAQ modules introduced at SK to insure the as much data as possible from a nearby supernova is recorded without overflowing or crashing the standard QBEE-based DAQ. 
These modules, the SN module and Veto module, effectively prescale PMT hits, vetoing the standard DAQ activity during periods with very high event rates to prevent buffer overflows.

The performance of these modules was checked with dedicated high-event-rate laser diode data taking and confirmed that the SN module takes dead-time free data and the Veto module prevents the QBEEs from overflowing. 
Further, an analysis of those data was successfully able to reconstruct the input number of diode flashes, indicating the robustness of the data taking during high rate scenarios. 
Long-term testing of the modules from June 2021 to May 2022 confirmed that the modules do not affect ordinary operations at SK and identified two types of events that trigger the Veto modules: energetic cosmic ray muons and flashing PMTs. 
However, even with those triggers less than 1\,ms of deadtime is expected per year with the new modules. 

Simulations were performed to understand the behavior of the modules in response to nearby supernovae using two different supernova models and assuming both pure and Gd-loaded detector conditions. 
Under all conditions we find that the modules will be triggered for a supernova located at or closer than 800\,pc.
For Betelgeuse, which is at 170\,pc away from the earth, the veto dead-time is found to be between 3\,s to 4\,s. 
In spite of this dead-time, the Veto module and SN module will allow us to nonetheless analyze the bulk of the supernova neutrino emission from Betelgeuse, which should continue for more than 1 minute.


\section*{Acknowlegements}

We gratefully acknowledge the cooperation of the Kamioka Mining and Smelting Company.
The SK experiment has been built and operated from funding by the 
Japanese Ministry of Education, Culture, Sports, Science and Technology, the U.S.
Department of Energy, and the U.S. National Science Foundation. Some of us have been 
supported by funds from the National Research Foundation of Korea (NRF-2009-0083526 
and NRF 2022R1A5A1030700) funded by the Ministry of Science, 
Information and Communication Technology (ICT), the Institute for 
Basic Science (IBS-R016-Y2), and the Ministry of Education (2018R1D1A1B07049158,
2021R1I1A1A01042256, the Japan Society for the Promotion of Science, the National
Natural Science Foundation of China under Grants No.11620101004, the Spanish Ministry of Science, 
Universities and Innovation (grant PGC2018-099388-B-I00), the Natural Sciences and 
Engineering Research Council (NSERC) of Canada, the Scinet and Westgrid consortia of
Compute Canada, the National Science Centre (UMO-2018/30/E/ST2/00441) and the Ministry
of Education and Science (DIR/WK/2017/05), Poland,
the Science and Technology Facilities Council (STFC) and
Grid for Particle Physics (GridPP), UK, the European Union's 
Horizon 2020 Research and Innovation Programme under the Marie Sklodowska-Curie grant
agreement no.754496, H2020-MSCA-RISE-2018 JENNIFER2 grant agreement no.822070, and 
H2020-MSCA-RISE-2019 SK2HK grant agreement no. 872549.

\bibliographystyle{ptephy}
\bibliography{sample}
\end{document}

%% file: authors-20221203.tex
\newcommand{\AFFicrr}{1}
\newcommand{\AFFkashiwa}{2}
\newcommand{\AFFicrronly}{3}
\newcommand{\AFFipmu}{4}
\newcommand{\AFFmad}{5}
\newcommand{\AFFubc}{6}
\newcommand{\AFFbu}{7}
\newcommand{\AFFuci}{8}
\newcommand{\AFFcsu}{9}
\newcommand{\AFFcnm}{10}
\newcommand{\AFFduke}{11}
\newcommand{\AFFfukuoka}{12}
\newcommand{\AFFgifu}{13}
\newcommand{\AFFgist}{14}
\newcommand{\AFFuh}{15}
\newcommand{\AFFicl}{16}
\newcommand{\AFFkek}{17}
\newcommand{\AFFkobe}{18}
\newcommand{\AFFkyoto}{19}
\newcommand{\AFFliv}{20}
\newcommand{\AFFmiyagi}{21}
\newcommand{\AFFnagoya}{22}
\newcommand{\AFFkmi}{23}
\newcommand{\AFFpol}{24}
\newcommand{\AFFsuny}{25}
\newcommand{\AFFokayama}{26}
\newcommand{\AFFosaka}{27}
\newcommand{\AFFox}{28}
\newcommand{\AFFqmul}{29}
\newcommand{\AFFregina}{30}
\newcommand{\AFFseoul}{31}
\newcommand{\AFFsheff}{32}
\newcommand{\AFFshizuokasc}{33}
\newcommand{\AFFstfc}{34}
\newcommand{\AFFskk}{35}
\newcommand{\AFFtodai}{36}
\newcommand{\AFFtit}{37}
\newcommand{\AFFtus}{38}
\newcommand{\AFFtoronto}{39}
\newcommand{\AFFtriumf}{40}
\newcommand{\AFFtokai}{41}
\newcommand{\AFFtsinghua}{42}
\newcommand{\AFFynu}{43}
\newcommand{\AFFllr}{44}
\newcommand{\AFFbari}{45}
\newcommand{\AFFnapoli}{46}
\newcommand{\AFFroma}{47}
\newcommand{\AFFpadova}{48}
\newcommand{\AFFkeio}{49}
\newcommand{\AFFwinnipeg}{50}
\newcommand{\AFFkcl}{51}
\newcommand{\AFFwarwick}{52}
\newcommand{\AFFral}{53}
\newcommand{\AFFwu}{54}
\newcommand{\AFFbcit}{55}
\newcommand{\AFFtohoku}{56}
\newcommand{\AFFicise}{57}
\newcommand{\AFFilance}{58}
\newcommand{\AFFibs}{59}

\author{
\name{M.~Mori}{\AFFkyoto},
\name{K.~Abe}{\AFFicrr,\AFFipmu},
\name{Y.~Hayato}{\AFFicrr,\AFFipmu},
\name{K.~Hiraide}{\AFFicrr,\AFFipmu},
\name{K.~Hosokawa}{\AFFicrr,\AFFipmu},
\name{K.~Ieki}{\AFFicrr,\AFFipmu},
\name{M.~Ikeda}{\AFFicrr,\AFFipmu},
\name{J.~Kameda}{\AFFicrr,\AFFipmu},
\name{Y.~Kanemura}{\AFFicrr},
\name{R.~Kaneshima}{\AFFicrr}
% ...
\collaborator{(The Super-Kamiokande Collaboration)}}

\affil[\AFFicrr]{{Kamioka Observatory, Institute for Cosmic Ray Research, University of Tokyo, Kamioka, Gifu 506-1205, Japan}}
\affil[\AFFkashiwa]{{Research Center for Cosmic Neutrinos, Institute for Cosmic Ray Research, University of Tokyo, Kashiwa, Chiba 277-8582, Japan}}
\affil[\AFFicrronly]{{Institute for Cosmic Ray Research, University of Tokyo, Kashiwa, Chiba 277-8582, Japan}}
\affil[\AFFipmu]{{Kavli Institute for the Physics and
Mathematics of the Universe (WPI), The University of Tokyo Institutes for Advanced Study,
University of Tokyo, Kashiwa, Chiba 277-8583, Japan }}
\affil[\AFFmad]{{Department of Theoretical Physics, University Autonoma Madrid, 28049 Madrid, Spain}}
\affil[\AFFubc]{{Department of Physics and Astronomy, University of British Columbia, Vancouver, BC, V6T1Z4, Canada}}
\affil[\AFFbu]{{Department of Physics, Boston University, Boston, MA 02215, USA}}
\affil[\AFFuci]{{Department of Physics and Astronomy, University of California, Irvine, Irvine, CA 92697-4575, USA }}
\affil[\AFFcsu]{{Department of Physics, California State University, Dominguez Hills, Carson, CA 90747, USA}}
\affil[\AFFcnm]{{Institute for Universe and Elementary Particles, Chonnam National University, Gwangju 61186, Korea}}
\affil[\AFFduke]{{Department of Physics, Duke University, Durham NC 27708, USA}}
\affil[\AFFfukuoka]{{Junior College, Fukuoka Institute of Technology, Fukuoka, Fukuoka 811-0295, Japan}}
\affil[\AFFgifu]{{Department of Physics, Gifu University, Gifu, Gifu 501-1193, Japan}}
\affil[\AFFgist]{{GIST College, Gwangju Institute of Science and Technology, Gwangju 500-712, Korea}}
\affil[\AFFuh]{{Department of Physics and Astronomy, University of Hawaii, Honolulu, HI 96822, USA}}
\affil[\AFFicl]{{Department of Physics, Imperial College London , London, SW7 2AZ, United Kingdom }}
\affil[\AFFkek]{{High Energy Accelerator Research Organization (KEK), Tsukuba, Ibaraki 305-0801, Japan }}
\affil[\AFFkobe]{{Department of Physics, Kobe University, Kobe, Hyogo 657-8501, Japan}}
\affil[\AFFkyoto]{{Department of Physics, Kyoto University, Kyoto, Kyoto 606-8502, Japan}}
\affil[\AFFliv]{{Department of Physics, University of Liverpool, Liverpool, L69 7ZE, United Kingdom}}
\affil[\AFFmiyagi]{{Department of Physics, Miyagi University of Education, Sendai, Miyagi 980-0845, Japan}}
\affil[\AFFnagoya]{{Institute for Space-Earth Environmental Research, Nagoya University, Nagoya, Aichi 464-8602, Japan}}
\affil[\AFFkmi]{{Kobayashi-Maskawa Institute for the Origin of Particles and the Universe, Nagoya University, Nagoya, Aichi 464-8602, Japan}}
\affil[\AFFpol]{{National Centre For Nuclear Research, 02-093 Warsaw, Poland}}
\affil[\AFFsuny]{{Department of Physics and Astronomy, State University of New York at Stony Brook, NY 11794-3800, USA}}
\affil[\AFFokayama]{{Department of Physics, Okayama University, Okayama, Okayama 700-8530, Japan }}
\affil[\AFFosaka]{{Department of Physics, Osaka University, Toyonaka, Osaka 560-0043, Japan}}
\affil[\AFFox]{{Department of Physics, Oxford University, Oxford, OX1 3PU, United Kingdom}}
\affil[\AFFqmul]{{School of Physics and Astronomy, Queen Mary University of London, London, E1 4NS, United Kingdom}}
\affil[\AFFregina]{{Department of Physics, University of Regina, 3737 Wascana Parkway, Regina, SK, S4SOA2, Canada}}
\affil[\AFFseoul]{{Department of Physics, Seoul National University, Seoul 151-742, Korea}}
\affil[\AFFsheff]{{Department of Physics and Astronomy, University of Sheffield, S3 7RH, Sheffield, United Kingdom}}
\affil[\AFFshizuokasc]{{Department of Informatics in
Social Welfare, Shizuoka University of Welfare, Yaizu, Shizuoka, 425-8611, Japan}}
\affil[\AFFstfc]{{STFC, Rutherford Appleton Laboratory, Harwell Oxford, and Daresbury Laboratory, Warrington, OX11 0QX, United Kingdom}}
\affil[\AFFskk]{{Department of Physics, Sungkyunkwan University, Suwon 440-746, Korea}}
\affil[\AFFtodai]{{Department of Physics, University of Tokyo, Bunkyo, Tokyo 113-0033, Japan }}
\affil[\AFFtit]{{Department of Physics,Tokyo Institute of Technology, Meguro, Tokyo 152-8551, Japan }}
\affil[\AFFtus]{{Department of Physics, Faculty of Science and Technology, Tokyo University of Science, Noda, Chiba 278-8510, Japan }}
\affil[\AFFtoronto]{{Department of Physics, University of Toronto, ON, M5S 1A7, Canada }}
\affil[\AFFtriumf]{{TRIUMF, 4004 Wesbrook Mall, Vancouver, BC, V6T2A3, Canada }}
\affil[\AFFtokai]{{Department of Physics, Tokai University, Hiratsuka, Kanagawa 259-1292, Japan}}
\affil[\AFFtsinghua]{{Department of Engineering Physics, Tsinghua University, Beijing, 100084, China}}
\affil[\AFFynu]{{Department of Physics, Yokohama National University, Yokohama, Kanagawa, 240-8501, Japan}}
\affil[\AFFllr]{{Ecole Polytechnique, IN2P3-CNRS, Laboratoire Leprince-Ringuet, F-91120 Palaiseau, France }}
\affil[\AFFbari]{{ Dipartimento Interuniversitario di Fisica, INFN Sezione di Bari and Universit\`a e Politecnico di Bari, I-70125, Bari, Italy}}
\affil[\AFFnapoli]{{Dipartimento di Fisica, INFN Sezione di Napoli and Universit\`a di Napoli, I-80126, Napoli, Italy}}
\affil[\AFFroma]{{INFN Sezione di Roma and Universit\`a di Roma ``La Sapienza'', I-00185, Roma, Italy}}
\affil[\AFFpadova]{{Dipartimento di Fisica, INFN Sezione di Padova and Universit\`a di Padova, I-35131, Padova, Italy}}
\affil[\AFFkeio]{{Department of Physics, Keio University, Yokohama, Kanagawa, 223-8522, Japan}}
\affil[\AFFwinnipeg]{{Department of Physics, University of Winnipeg, MB R3J 3L8, Canada }}
\affil[\AFFkcl]{{Department of Physics, King's College London, London, WC2R 2LS, UK }}
\affil[\AFFwarwick]{{Department of Physics, University of Warwick, Coventry, CV4 7AL, UK }}
\affil[\AFFral]{{Rutherford Appleton Laboratory, Harwell, Oxford, OX11 0QX, UK }}
\affil[\AFFwu]{{Faculty of Physics, University of Warsaw, Warsaw, 02-093, Poland }}
\affil[\AFFbcit]{{Department of Physics, British Columbia Institute of Technology, Burnaby, BC, V5G 3H2, Canada }}
\affil[\AFFtohoku]{{Department of Physics, Faculty of Science, Tohoku University, Sendai, Miyagi, 980-8578, Japan }}
\affil[\AFFicise]{{Institute For Interdisciplinary Research in Science and Education, ICISE, Quy Nhon, 55121, Vietnam }}
\affil[\AFFilance]{{ILANCE, CNRS - University of Tokyo International Research Laboratory, Kashiwa, Chiba 277-8582, Japan}}
\affil[\AFFibs]{{Institute for Basic Science (IBS), Daejeon, 34126, Korea}}